\documentclass[%
 reprint,
superscriptaddress,
 amsmath,amssymb,
 aps,
prc,
]{revtex4-1}

\usepackage{graphicx}
\usepackage{dcolumn}
\usepackage{bm}
\usepackage[modulo]{lineno}

\usepackage{float}

\usepackage{booktabs}
\newcolumntype{C}[1]{>{\centering\arraybackslash}p{#1}}
\AtBeginDocument{
\heavyrulewidth=.08em
\lightrulewidth=.05em
\cmidrulewidth=.03em
\belowrulesep=.65ex
\belowbottomsep=0pt
\aboverulesep=.4ex
\abovetopsep=0pt
\cmidrulesep=\doublerulesep
\cmidrulekern=.5em
\defaultaddspace=.5em
}

\begin{document}

\preprint{APS/123-QED}

\title{Total absorption $\gamma$-ray spectroscopy of niobium isomers}

\author{V. Guadilla}
\email{guadilla@ific.uv.es}
\altaffiliation[\newline Present address: ]{Subatech, IMT-Atlantique, Universit\'e de Nantes, CNRS-IN2P3, F-44307, Nantes, France}
\affiliation{%
 Instituto de F\'isica Corpuscular, CSIC-Universidad de Valencia, E-46071, Valencia, Spain
}
\author{A. Algora}%
\email{algora@ific.uv.es}
\affiliation{%
 Instituto de F\'isica Corpuscular, CSIC-Universidad de Valencia, E-46071, Valencia, Spain
}
\affiliation{%
 Institute of Nuclear Research of the Hungarian Academy of Sciences, Debrecen H-4026, Hungary.
}
\author{J. L. Tain}%
\affiliation{%
 Instituto de F\'isica Corpuscular, CSIC-Universidad de Valencia, E-46071, Valencia, Spain
}
\author{J. Agramunt}  
\affiliation{%
 Instituto de F\'isica Corpuscular, CSIC-Universidad de Valencia, E-46071, Valencia, Spain
}
\author{J. \"Ayst\"o}  
\affiliation{%
 University of Jyv\"askyl\"a, FIN-40014, Jyv\"askyl\"a, Finland
}
\author{J. A. Briz}  
\affiliation{%
 Subatech, IMT-Atlantique, Universit\'e de Nantes, CNRS-IN2P3, F-44307, Nantes, France
}
\author{A. Cucoanes}  
\affiliation{%
 Subatech, IMT-Atlantique, Universit\'e de Nantes, CNRS-IN2P3, F-44307, Nantes, France
}
\author{T. Eronen}  
\affiliation{%
 University of Jyv\"askyl\"a, FIN-40014, Jyv\"askyl\"a, Finland
}
\author{M. Estienne}  
\affiliation{%
 Subatech, IMT-Atlantique, Universit\'e de Nantes, CNRS-IN2P3, F-44307, Nantes, France
}
\author{M. Fallot}  
\affiliation{%
 Subatech, IMT-Atlantique, Universit\'e de Nantes, CNRS-IN2P3, F-44307, Nantes, France
}
\author{L. M. Fraile}  
\affiliation{%
Grupo de F\'isica Nuclear and IPARCOS, Universidad Complutense de Madrid, CEI Moncloa, E-28040 Madrid, Spain
}
\author{E. Ganio\u{g}lu}  
\affiliation{%
Department of Physics, Istanbul University, 34134, Istanbul, Turkey
}
\author{W. Gelletly}  
\affiliation{%
Department of Physics, University of Surrey, GU2 7XH, Guildford, UK
} 
\author{D. Gorelov}  
\author{J. Hakala} 
\author{A. Jokinen}
\affiliation{%
 University of Jyv\"askyl\"a, FIN-40014, Jyv\"askyl\"a, Finland
}\author{D. Jordan} 
\affiliation{%
 Instituto de F\'isica Corpuscular, CSIC-Universidad de Valencia, E-46071, Valencia, Spain
}  
\author{A. Kankainen}  
\author{V. Kolhinen}  
\author{J. Koponen}  
\affiliation{%
 University of Jyv\"askyl\"a, FIN-40014, Jyv\"askyl\"a, Finland
}
\author{M. Lebois}  
\affiliation{%
Institut de Physique Nucl\`eaire d'Orsay, 91406, Orsay, France
}
\author{L. Le Meur}  
\affiliation{%
 Subatech, IMT-Atlantique, Universit\'e de Nantes, CNRS-IN2P3, F-44307, Nantes, France
}
\author{T. Martinez}  
\affiliation{%
Centro de Investigaciones Energ\'eticas Medioambientales y Tecnol\'ogicas, E-28040, Madrid, Spain
}
\author{M. Monserrate}  
\author{A. Montaner-Piz\'a}  
\affiliation{%
 Instituto de F\'isica Corpuscular, CSIC-Universidad de Valencia, E-46071, Valencia, Spain
}
\author{I. Moore}  
\affiliation{%
 University of Jyv\"askyl\"a, FIN-40014, Jyv\"askyl\"a, Finland
}
\author{E. N\'acher}  
\affiliation{%
Instituto de Estructura de la Materia, CSIC, E-28006, Madrid, Spain
}
\author{S. E. A. Orrigo}  
\affiliation{%
 Instituto de F\'isica Corpuscular, CSIC-Universidad de Valencia, E-46071, Valencia, Spain
}
\author{H. Penttil\"a}  
\author{I. Pohjalainen}  
\affiliation{%
 University of Jyv\"askyl\"a, FIN-40014, Jyv\"askyl\"a, Finland
}
\author{A. Porta}  
\affiliation{%
 Subatech, IMT-Atlantique, Universit\'e de Nantes, CNRS-IN2P3, F-44307, Nantes, France
}
\author{J. Reinikainen}  
\author{M. Reponen}  
\author{S. Rinta-Antila}  
\affiliation{%
 University of Jyv\"askyl\"a, FIN-40014, Jyv\"askyl\"a, Finland
}
\author{B. Rubio}  
\affiliation{%
 Instituto de F\'isica Corpuscular, CSIC-Universidad de Valencia, E-46071, Valencia, Spain
}
\author{K. Rytk\"onen}  
\affiliation{%
 University of Jyv\"askyl\"a, FIN-40014, Jyv\"askyl\"a, Finland
}
\author{P. Sarriguren}  
\affiliation{%
Instituto de Estructura de la Materia, CSIC, E-28006, Madrid, Spain
}
\author{T. Shiba}  
\affiliation{%
 Subatech, IMT-Atlantique, Universit\'e de Nantes, CNRS-IN2P3, F-44307, Nantes, France
}
\author{V. Sonnenschein}  
\affiliation{%
 University of Jyv\"askyl\"a, FIN-40014, Jyv\"askyl\"a, Finland
}
\author{A. A. Sonzogni}  
\affiliation{%
NNDC, Brookhaven National Laboratory, Upton, NY 11973-5000, USA
}
\author{E. Valencia}  
\affiliation{%
 Instituto de F\'isica Corpuscular, CSIC-Universidad de Valencia, E-46071, Valencia, Spain
}
\author{V. Vedia}  
\affiliation{%
Grupo de F\'isica Nuclear and IPARCOS, Universidad Complutense de Madrid, CEI Moncloa, E-28040 Madrid, Spain
}
\author{A. Voss} 
\affiliation{%
 University of Jyv\"askyl\"a, FIN-40014, Jyv\"askyl\"a, Finland
}
\author{J. N. Wilson}
\affiliation{%
Institut de Physique Nucl\`eaire d'Orsay, 91406, Orsay, France
}
\author{A. -A. Zakari-Issoufou} 
\affiliation{%
 Subatech, IMT-Atlantique, Universit\'e de Nantes, CNRS-IN2P3, F-44307, Nantes, France
}

\date{\today}

\begin{abstract}
The $\beta$ intensity distributions of the decays of $^{100\text{gs},100\text{m}}$Nb and $^{102\text{gs},102\text{m}}$Nb have been determined using the Total Absorption  $\gamma$-Ray Spectroscopy technique. The JYFLTRAP double Penning trap system was employed to disentangle the isomeric states involved, lying very close in energy, in a campaign of challenging measurements performed with the Decay Total Absorption $\gamma$-ray Spectrometer at the Ion Guide Isotope Separator On-Line facility in Jyv\"askyl\"a. The low-spin isomeric state of each niobium case was populated through the decay of the zirconium parent, that was treated as a contaminant. We have applied a method to extract this contamination, and additionally we have obtained $\beta$ intensity distributions for these zirconium decays. The $\beta$-strength distributions evaluated with these results were compared with calculations in quasiparticle random-phase approximation, suggesting a prolate configuration for the ground states of $^{100,102}$Zr. The footprint of the Pandemonium effect was found when comparing our results for the analyses of the niobium isotopes with previous decay data. The $\beta$-intensities of the decay of $^{102\text{m}}$Nb were obtained for the first time. A careful evaluation of the uncertainties was carried out, and the consistency of our results was validated taking advantage of the segmentation of our spectrometer. The final results were used as input in reactor summation calculations. A large impact on antineutrino spectrum calculations was already reported and here we detail the significant impact on decay heat calculations. 
\end{abstract}

\maketitle

\section{Introduction}

Every fission occurring in the nuclear fuel of a reactor is followed on average by six $\beta$ decays. This points to the relevance of the $\beta$ decay process in reactor applications. As a consequence, nuclear power reactors are the largest man-made non-military sources of antineutrinos, which can be used for antineutrino oscillation experiments \cite{DoubleChooz,DayaBay,Reno} and for reactor monitoring~\cite{non_proliferation_2017}. The $\beta$ decay of fission products is also responsible for approximately 7-8$\%$ of the energy released in a working nuclear reactor. This fraction of the released energy, named decay heat (DH),  becomes the dominant source of energy after reactor shutdown because of the wide range of half-lives in the decay of fission products. DH calculations are therefore a basic need for the design and safe manipulation of nuclear reactors, as well as for the storage and transport of the nuclear waste. These calculations require large databases with reliable nuclear data. $\beta$ decay half-lives and the released mean $\gamma$ and $\beta$ energies per $\beta$ decay are the essential ingredients needed from these databases (see for example \cite{JEFF}). 

The mean $\gamma$ and $\beta$ energies per $\beta$ decay can be obtained from direct measurements \cite{Rudstam, Olof_beta} or can be calculated from the known $\beta$ decay level schemes available in evaluated nuclear structure databases like ENSDF \cite{ENSDF}. Incomplete $\beta$ decay schemes lead to incorrect determination of the mean energies per decay. This incompleteness can be due to the limitations in efficiency of HPGe detectors used in high-resolution spectroscopy studies of $\beta$ decay. The problem arises from the incorrect determination of the $\beta$ transition probabilities. In high-resolution spectroscopy experiments $\beta$ decay probabilities are obtained from the $\gamma$ intensity balance between the population and the de-excitation of the levels fed in $\beta$ decay. In complex $\beta$ decays, levels populated at high excitation energy in the daughter nucleus can de-excite through many possible weak $\gamma$ decay branches, and some of those branches occur through the emission of $\gamma$-rays of relatively high energy. The detection of high-energy $\gamma$ transitions, as well as the detection of weak $\gamma$-rays in general, is not always possible with the limited efficiency of HPGe detectors. Even with HPGe arrays of high efficiency, $\beta$ feeding at high excitation in the daughter nucleus may remain undetermined \cite{Algora_Ho}. This experimental difficulty is known in the literature as the Pandemonium effect \cite{Pandemonium}. It can be avoided using the Total Absorption $\gamma$-ray Spectroscopy (TAGS) technique, which relies on the use of $\gamma$-ray calorimeters to detect the full $\gamma$ cascades that follow the $\beta$ decay~\cite{Algora_npn,Rubio_tas}. The potential of using the TAGS technique in $\beta$ decay measurements important for reactor calculations has already been demonstrated \cite{DecayHeat,LoliTc,neutrinos_PRL,Zak_PRL,MTAS_neutrino_PRL,vTAS_PRC,Simon_PRC,MTAS_neutrino_DH_PRL}.
 
We present here the TAGS study of the $\beta$ decays of $^{100\text{gs},100\text{m}}$Nb and $^{102\text{gs},102\text{m}}$Nb. In the context of reactor applications, the $\beta$ decay of Nb isotopes with $A \sim 100$ is considered of high relevance. The decay of the $^{100\text{gs}}$Nb was identified as a case that should be measured using TAGS with the highest priority in relation to the DH of U/Pu fuel~\cite{NEA_IAEA_DecayHeat1}. Similarly the decays of $^{100\text{m}, 102\text{m}}$Nb were identified as first priority decays for the Th/U fuel~\cite{NEA_IAEA_DecayHeat2}. Their relevance can be understood in terms of their relatively high cumulative fission yields in the fission of U/Pu and Th/U fuels. These decays are also considered to be of high interest in the framework of antineutrino summation calculations \cite{Sonzogni_summation,Zak_PRL} that have recently attracted considerable attention in relation to the reactor anomaly and the reactor antineutrino spectrum distortion \cite{Anomaly,RENO_shoulder,DayaBay_shoulder,DoubleChooz_shoulder}. 

This article is a follow-up to our recent work \cite{PRL_Nb}, where we emphasized the relevance of the decays studied in antineutrino spectrum summation calculations. The $\beta$-decaying ground states and isomeric states present in $^{100,102}$Nb are very close in energy and they have similar decay half-lives. For this reason, the disentanglement of the decaying isomers required the use of different strategies. In the present publication we provide more details of the production strategies and on the analyses of the decays of interest, that were not covered in \cite{PRL_Nb}. We also give details of the impact on DH summation calculations. Nuclear structure aspects of the $\beta$ decay of $^{100,102}$Zr will also be discussed. The $\beta$ decay of $^{100,102}$Zr was used for the selective production of the low-spin isomers in $^{100,102}$Nb, since the ground states of the even-even Zr isotopes have $J^\pi=0^{+}$, and their $\beta$ decays do not populate the high-spin isomeric states in the corresponding Nb nuclei. The $\beta$ strengths deduced for the decays of $^{100,102}$Zr will be compared with QRPA calculations to infer the shapes of the ground states of the Zr isotopes, following the line of previous works~\cite{KikeShape,PRC_Poirier,PRC_AnaBelen,PRC_Esther,PRC_Briz}. 

The structure of the article is as follows: in section \ref{exp} we will present details of the experiment, in section \ref{Zr} the analyses of the $\beta$ decay of $^{100,102}$Zr will be discussed, and in sections \ref{100Nb} and \ref{102Nb} the TAGS analyses of the decays of the $^{100,102}$Nb low and high-spin isomers will be presented. Different checks on the analysis will be discussed in section \ref{crosschecks}, exploiting the segmentation of our spectrometer. Finally, in section \ref{reactor} the impact of the measurements on DH summation calculations will be discussed. Conclusions will be drawn in section \ref{conclusions}.

\section{Experiment}\label{exp}

\begin{figure}[h]
\begin{center}
\includegraphics[width=0.5 \textwidth]{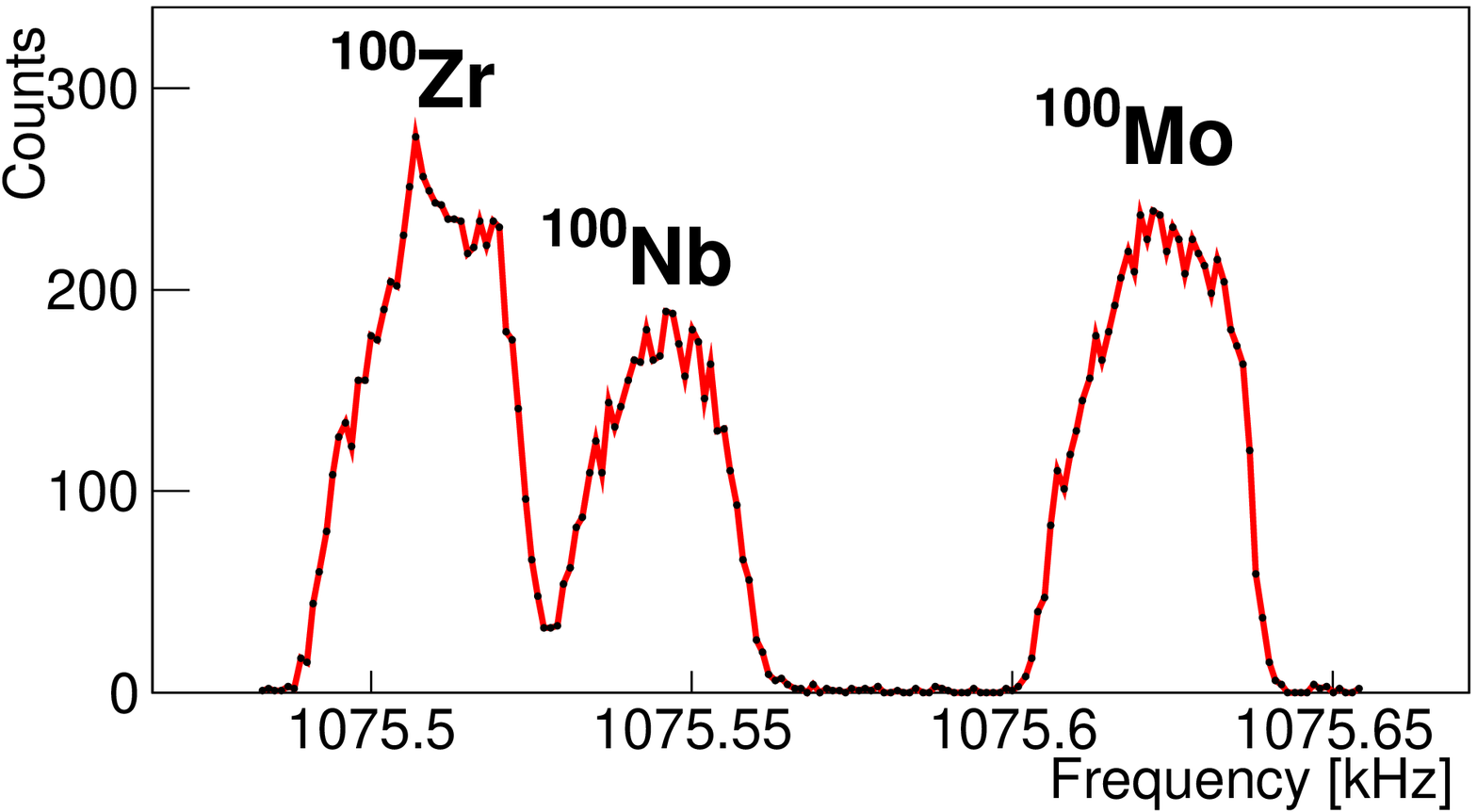}

\includegraphics[width=0.5 \textwidth]{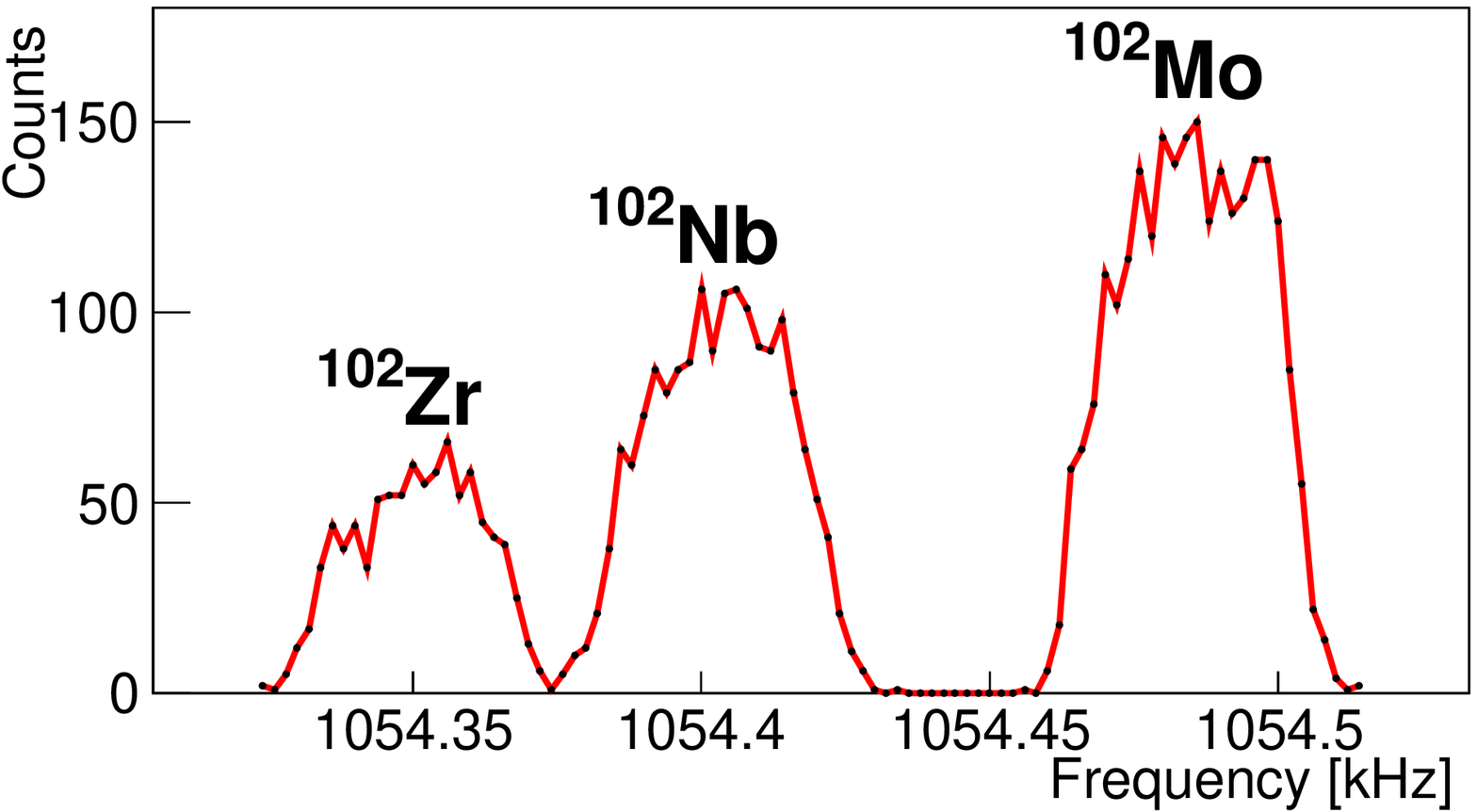}
\caption{JYFLTRAP purification trap mass scans for A=100 and A=102. The corresponding frequency is selected to extract the nuclei of interest from the trap.}
\label{Nb_massScan} 
\end{center}
\end{figure}

The nuclei of interest were produced at the Ion Guide Isotope Separator On-Line (IGISOL) facility~\cite{Moore_IGISOLIV} in Jyv\"askyl\"a by proton-induced fission on a natural uranium target with 25~MeV protons from the K130 cyclotron. The fission ion guide technique~\cite{ION_Guide_1, ION_Guide_2} employed allows the extraction of refractory elements, such as the niobium and zirconium cases studied here. Approximately 1$\%$ of the resulting fission fragments are stopped in 300~mbar helium buffer gas with the majority remaining singly-charged. Following extraction from the gas cell through a differential pumping system and sextupole ion guide SPIG~\cite{Karvonen2008}, ions are accelerated to 30~keV and initially mass-separated using a 55 degree dipole magnet with a modest resolving power (M/$\Delta \text{M}\simeq$ 300-500). After cooling and bunching in the radio-frequency quadrupole cooler and buncher~\cite{IGISOL_cooling}, the JYFLTRAP double Penning trap system~\cite{JYFLTRAP} allows for a high-resolution mass separation. The clean separation of the isobars obtained with JYFLTRAP for the cases studied in this work is presented in Fig.~\ref{Nb_massScan}. The radioactive beam was delivered to the experimental set-up consisting of the Decay Total Absorption $\gamma$-ray Spectrometer (DTAS)~\cite{DTAS_design}, composed of 18 NaI(Tl) crystals, a plastic $\beta$ detector of 3~mm thickness located at the centre of DTAS, and a HPGe detector placed behind the $\beta$ detector~\cite{NIMB_DTAS}. The beam was implanted on a moving tape placed in front of the $\beta$ detector. The implantation and transport cycles of the tape system were selected according to the half-life of the nuclei studied in order to reduce the impact of the daughter activity in the measurements. 

\begin{figure}[h]
\begin{center}
\includegraphics[width=0.5 \textwidth]{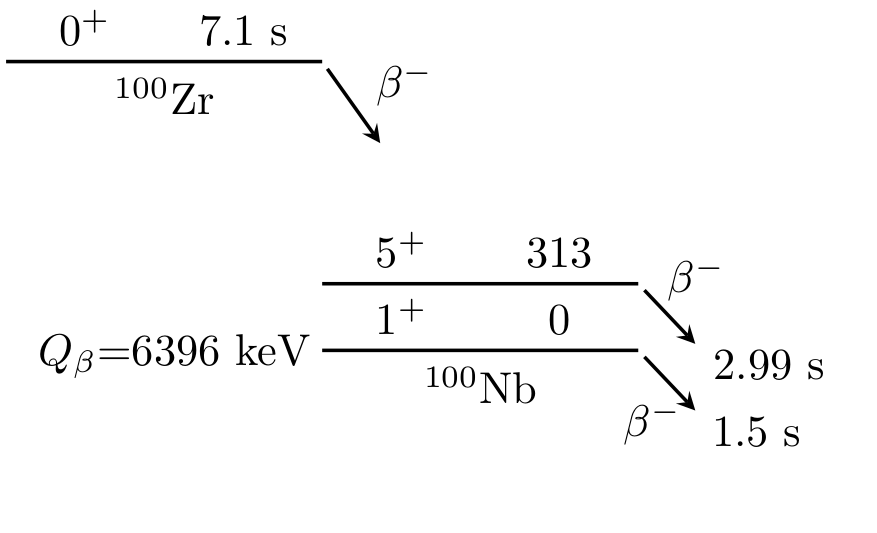}

\includegraphics[width=0.5 \textwidth]{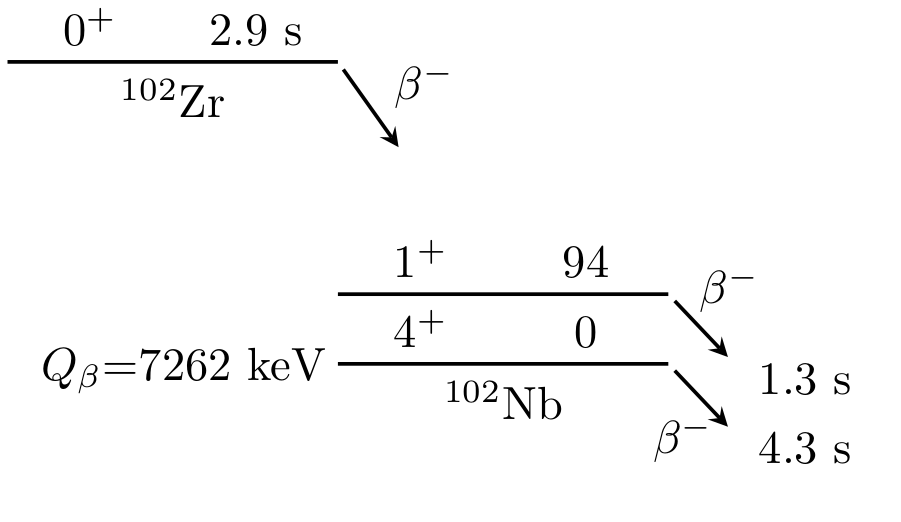}
\caption{Decay schemes for the $^{100}$Nb and $^{102}$Nb systems. Spin-parities of the $\beta$-decaying states involved and their half-lives are presented (values from ENSDF~\cite{NDS_A100,NDS_A102}). The $Q_{\beta}$ of each ground state (from the Atomic Mass Evaluation (AME) 2016~\cite{AME2016}) and the energies (in keV) of both isomeric states (from NUBASE 2016~\cite{NUBASE2016}) are indicated. The decay of the ground state of the parents, $^{100,102}$Zr, is also shown.}
\label{Nb_scheme} 
\end{center}
\end{figure}

In Fig.~\ref{Nb_scheme} we present the main $\beta$-decay properties of the A=100 and A=102 systems studied in this work. The existence of metastable states at low excitation energies (313 and 94~keV in $^{100}$Nb and $^{102}$Nb, respectively~\cite{NUBASE2016}) makes the measurements especially challenging. In addition, as can be seen in Fig.~\ref{Nb_scheme}, the half-life difference between the pairs of decaying states is small~\cite{NDS_A100, NDS_A102}, making the separation by means of different implantation-measuring cycles also difficult. In our study we followed similar strategies in both cases in order to distinguish experimentally between each pair of decaying isomers:

\begin{itemize}
\item The low-spin isomers were populated selectively through the $\beta$ decay of the 0$^{+}$ ground states of the zirconium parents. The zirconium isotopes were extracted from JYFLTRAP by selecting the corresponding frequency associated with the well-separated peaks shown in Fig.~\ref{Nb_massScan}. From these first measurements we were able to analyze the low-spin isomers by considering the decay of the zirconium parents as a contamination. No contamination from the high-spin isomer decays is found. After the evaluation of the Zr contamination, from these measurements clean low-spin Nb isomer decay spectra were obtained and analyzed. 

\item A combined measurement of both  isomers was performed for each case, selecting in the trap the corresponding frequency for the niobium peak shown in Fig.~\ref{Nb_massScan}. In this case a mixture of high- and low-spin isomers was obtained. From these measurements we were able to analyze the high-spin isomers, for they were favoured in the proton-induced fission process. The low-spin component was treated as a contaminant determined from the clean low-spin spectra obtained from the Zr measurements.

\end{itemize}

\begin{figure}[h]
\begin{center}
\includegraphics[width=0.5 \textwidth]{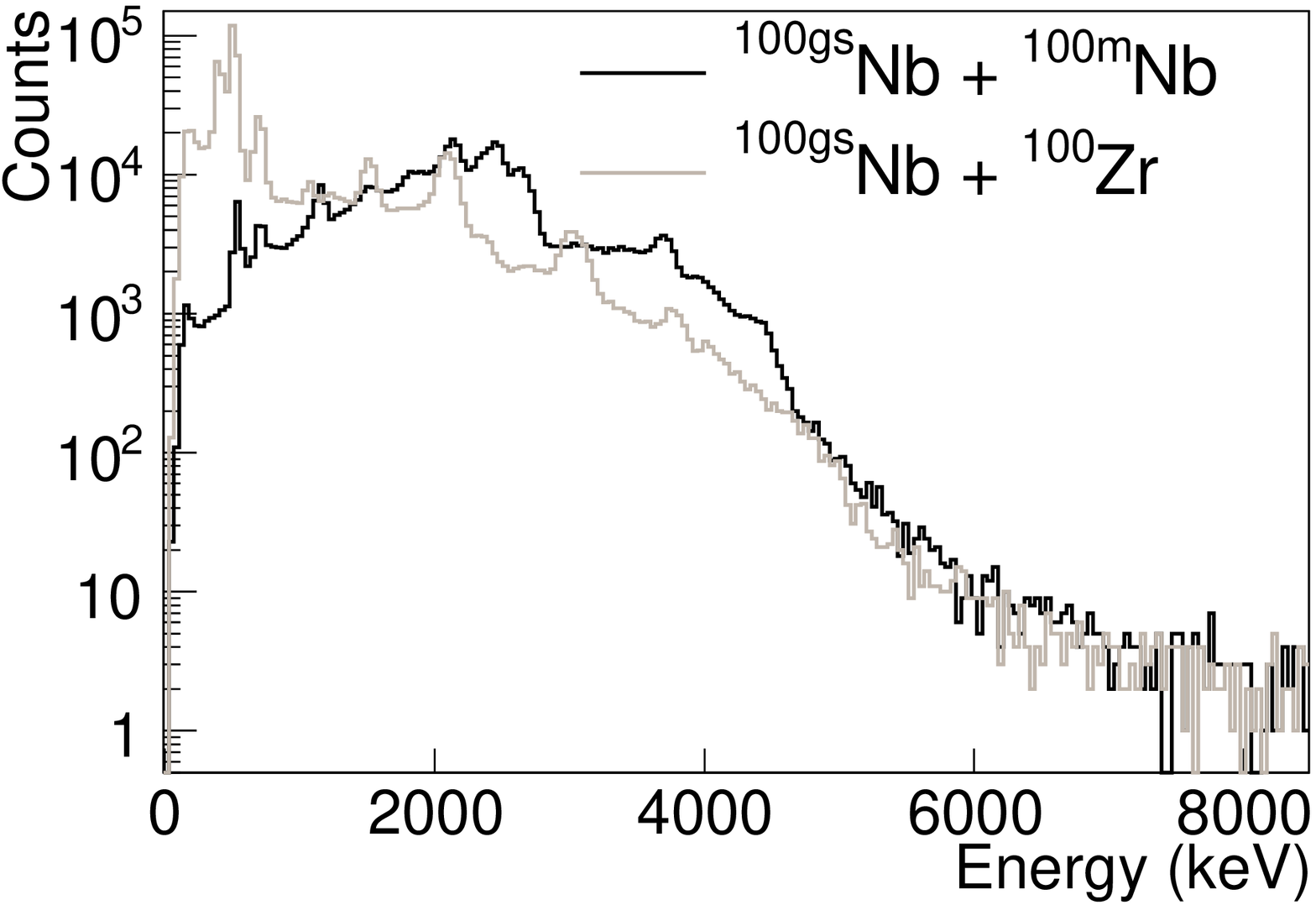}

\includegraphics[width=0.5 \textwidth]{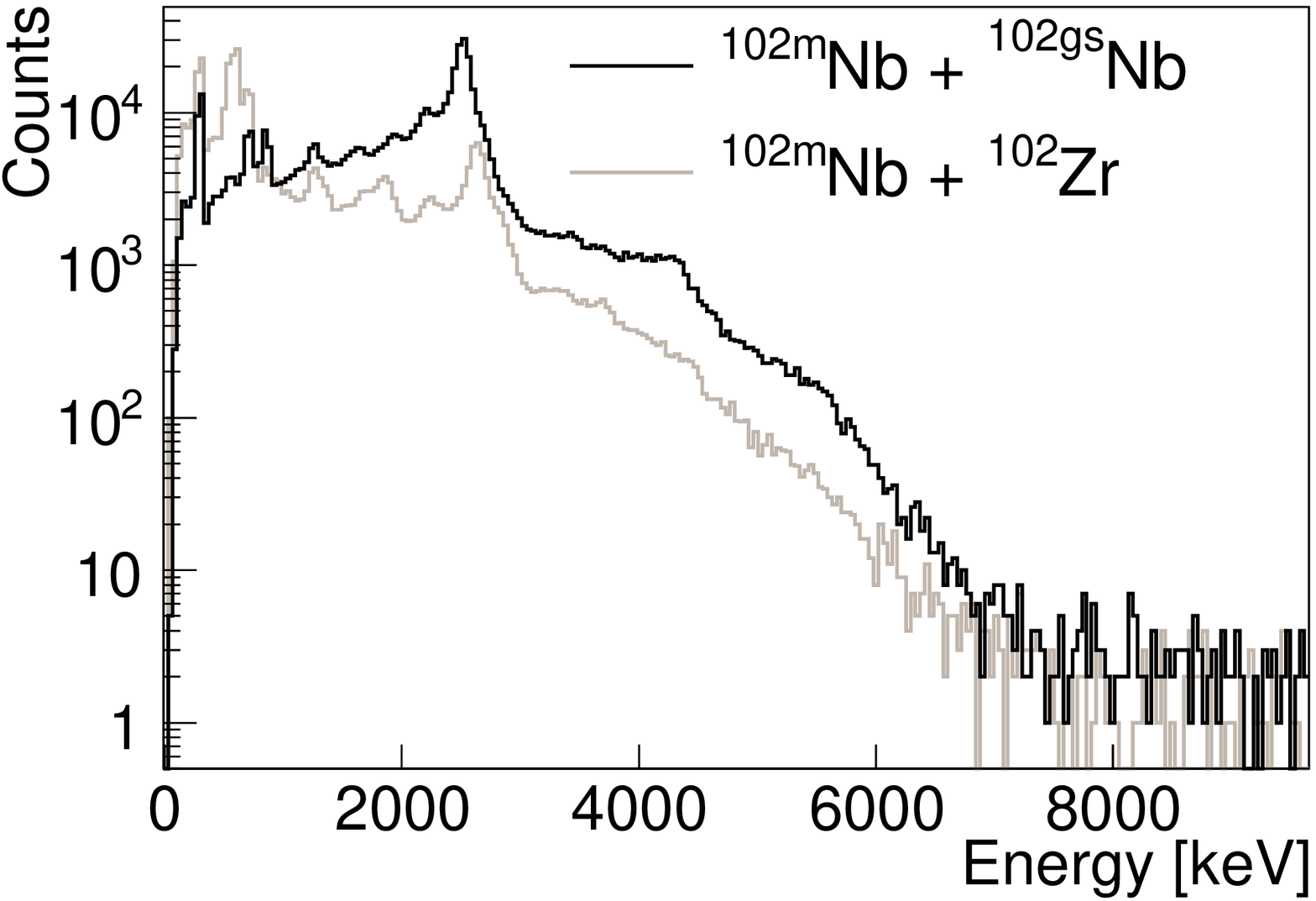}
\caption{Experimental $\beta$-gated spectra of the measurements for A=100 (top) and A=102 (bottom). The combined measurement of the decays of the low-spin isomer and the zirconium parent (gray) and the combined measurement of the decays of the two isomers (black) are shown for both cases.}
\label{Nb_measurements} 
\end{center}
\end{figure}

In the measurements a coincidence between DTAS and the $\beta$-detector ($\beta$-gated spectrum) allowed us to reject the environmental background. The total energy sum of the eighteen crystals of DTAS was reconstructed following the procedure described in \cite{NIMA_DTAS}. The $\beta$-gated spectra for the set of measurements described above is shown in Fig.~\ref{Nb_measurements}. The summing-pileup distortion, present in all our measurements, was calculated as in previous works \cite{vTAS_PRC,Simon_PRC,Zak_PRL,100Tc} with a Monte Carlo (MC) procedure based on the random superposition of two stored events within the ADC gate \cite{TAS_pileup,NIMA_DTAS}.

The TAGS analyses of the experimental $\beta$-gated spectra were performed following the method developed by the Valencia group \cite{TAS_algorithms,TAS_decaygen} to solve the inverse problem represented by:

\begin{center}
\begin{equation}\label{inverse}
d_i=\sum\limits_{j}R_{ij}(B)f_j+C_i
\end{equation}
\end{center}

\noindent with $d_i$ the number of counts in channel $i$ of the measured spectrum, $f_j$ the number of events that fed level $j$ in the daughter nucleus, $C_i$ the contribution of all contaminants to channel $i$, and  $R_{ij}(B)$ the response function of the detector. $R_{ij}$ represents the probability that feeding to level $j$ gives a count in channel $i$ of the experimental spectrum. $R_{ij}$ is unique for each detector set-up and depends on the de-exciting branching-ratio matrix (B) of the levels in the daughter nucleus~\cite{TAS_algorithms}. The branching-ratio matrix is calculated by combining the known decay information of the levels at low excitation energies available from the literature with the statistical model at higher energies. From the last known level included from high resolution measurements up to $Q_{\beta}$ we define a continuum region with 40 keV bins and we determine their $\gamma$ decay branching ratios based on a statistical model \cite{TAS_decaygen}. This model uses level densities and $\gamma$ strength functions that we take from the RIPL-3 \cite{RIPL-3} database. The parameters used for the cases studied in this work are summarized in Table \ref{parameters}. The quadrupole deformation parameter ($\beta$) and the level density parameter ``a'' at the neutron binding energy are needed for the parametrization of the E1 $\gamma$ strength function. For $^{100,102}$Nb the parameter $\beta$ was taken from experimental measurements \cite{DeformationPar_exp}, while for $^{100,102}$Mo it was taken from the Finite Range Droplet Model (FRDM) calculations available at RIPL-3. The parameter ``a'' was retrieved from the TALYS nuclear reaction code \cite{TALYS}. For the level density parametrization, the Hartree-Fock-Bogoliubov (HFB) plus combinatorial nuclear level densities \cite{Gorieli1,Gorieli2} available at RIPL-3 have been used. The C and P correction parameters from RIPL-3 were used when available. In addition, for $^{102}$Mo we have also tried an alternative P correction parameter that reproduces better the experimental number of levels at low energies (0.4 instead of the original value of 1.17951). Once the branching ratio matrix for a decay is constructed, the response function is calculated with previously validated MC simulations. The GEANT4 simulation package \cite{GEANT4} is used for this purpose. The simulation code was validated for this experiment by comparing MC simulations with measurements for a set of calibration sources ($^{22,24}$Na, $^{60}$Co, $^{137}$Cs and a mixed $^{152}$Eu-$^{133}$Ba source)~\cite{NIMA_DTAS}.

\begin{table*}[!h]
\begin{ruledtabular}
\caption{\label{parameters} Parameters used in the statistical model calculations of the branching ratio matrices (B) of the daughter nuclei.}
  \begin{tabular}{@{}C{1.cm}C{2.5cm}C{2cm}C{1cm}C{1cm}C{1cm}C{1cm}C{1cm}C{1cm}C{1cm}C{1cm}C{1cm}@{}}
    Nucleus
    & Level-density parameter
    & Deformation parameter
    & \multicolumn{9}{c}{Photon strength function parameters}
    \\ \cmidrule(lr){4-12}
    & &  & \multicolumn{3}{c}{E1} & \multicolumn{3}{c}{M1} & \multicolumn{3}{c}{E2}   \\
    \cmidrule(r){2-2}\cmidrule(lr){3-3}\cmidrule(lr){4-6}\cmidrule(lr){7-9}\cmidrule(l){10-12}
     & a($S_n$) & $\beta$  & E & $\Gamma$ & $\sigma$ & E & $\Gamma$ & $\sigma$ & E & $\Gamma$ & $\sigma$  \\
    & $[$MeV$^{-1}$] &  & [MeV] & [MeV] & [mb] & [MeV] & [MeV] & [mb] & [MeV] & [MeV] & [mb]  \\
    \\ \cmidrule(lr){1-12}   
    $^{100}$Nb & 18.168 &  0.358
    & \begin{tabular}{@{}c@{}}13.610\\ 18.344\end{tabular} 
    & \begin{tabular}{@{}c@{}}3.709\\ 6.541\end{tabular} 
    & \begin{tabular}{@{}c@{}}87.187 \\ 98.891\end{tabular} 
    & 8.847 & 4.000 & 0.639 & 13.594 & 4.901 & 2.048 
    \\ \cmidrule(lr){1-12}
    $^{100}$Mo & 16.594 &  0.231
    & \begin{tabular}{@{}c@{}}14.418\\ 17.513\end{tabular} 
    & \begin{tabular}{@{}c@{}}4.139\\ 5.989\end{tabular} 
    & \begin{tabular}{@{}c@{}}78.680 \\ 108.750\end{tabular} 
    & 8.847 & 4.000 & 0.850 & 13.594 & 4.910 & 2.149 
    \\ \cmidrule(lr){1-12}
    $^{102}$Nb & 17.600 &  0.378
    & \begin{tabular}{@{}c@{}}13.431\\ 18.400\end{tabular} 
    & \begin{tabular}{@{}c@{}}3.617\\ 6.578\end{tabular} 
    & \begin{tabular}{@{}c@{}}90.620\\ 99.662\end{tabular} 
    & 8.789 & 4.000 & 0.634 & 13.504 & 4.886 & 2.017  
    \\ \cmidrule(lr){1-12}
    $^{102}$Mo & 17.680 &  0.311
    & \begin{tabular}{@{}c@{}}13.829\\ 17.939\end{tabular} 
    & \begin{tabular}{@{}c@{}}3.824\\ 6.269\end{tabular} 
    & \begin{tabular}{@{}c@{}}86.385\\ 105.375\end{tabular} 
    & 8.789 & 4.000 & 0.879 & 13.504 & 4.886 & 2.117       
  \end{tabular}
\end{ruledtabular}
\end{table*}

\section{Analyses of $^{100,102}$Z\lowercase{r}}\label{Zr}

The decays of $^{100}$Zr and $^{102}$Zr were used to produce the low-spin niobium isomers, and hence they are considered as contaminants in the analyses of the $^{100\text{gs}}$Nb and $^{102\text{m}}$Nb decays, as explained above. First we need to assess how well this contamination is accounted for in our data. In the literature, we find information for both zirconium decays coming from high-resolution measurements also performed at IGISOL~\cite{Zr_Sami}. The $Q_{\beta}$ of the decay of $^{100}$Zr is 3.421~MeV and the last level populated in $\beta$-decay that was seen experimentally is at 704.1~keV, while for the decay of $^{102}$Zr $Q_{\beta}$=4.717~MeV and the last level populated in $\beta$-decay seen experimentally is at 940.5~keV \cite{Zr_Sami}. This information seems incomplete, and in fact the authors of \cite{Zr_Sami} suggest that part of the $\beta$-intensity could have remained unobserved. In both cases a TAGS measurement would be advisable to study the potentially missing $\beta$ intensity. In our experimental campaign, however, we did not have time to accomplish such a goal, that would have required several measurements with different implantation cycles due to the closeness of the half-lives of the zirconium and niobium decays. In spite of this lack of experimental measurements, we have exploited the time information registered within each measurement cycle to sort the TAGS data off-line using different time windows to disentangle them. 

\begin{figure}[h]
\begin{center}
\includegraphics[width=0.5 \textwidth]{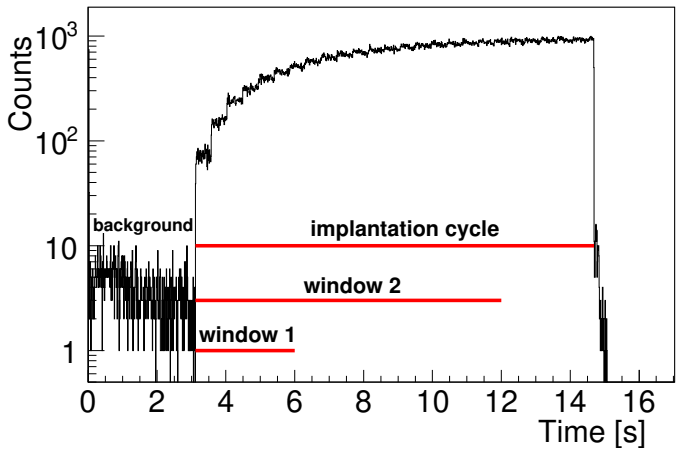}
\caption{Number of $\beta$ particles detected in the plastic detector versus time in the cycle for the combined measurement of the decays of $^{102}$Zr and $^{102\text{m}}$Nb. The first part of the spectrum corresponds to no implantation, useful to check the level of background previous to the measurement. Two different sorting time windows are shown as an example.}
\label{Grow} 
\end{center}
\end{figure}

The method employed here relies on the possibility of setting different time windows offline for the implantation cycle up to the maximum cycle length fixed in the experimental measurement. This is possible thanks to a clock signal registered in our ADC and it allows us to study the number of $\beta$ particles detected in the plastic detector as a function of time, as shown in Fig.~\ref{Grow} for the $^{102}$Zr+$^{102\text{m}}$Nb measurement.

\begin{figure}[h]
\begin{center}
\includegraphics[width=0.5 \textwidth]{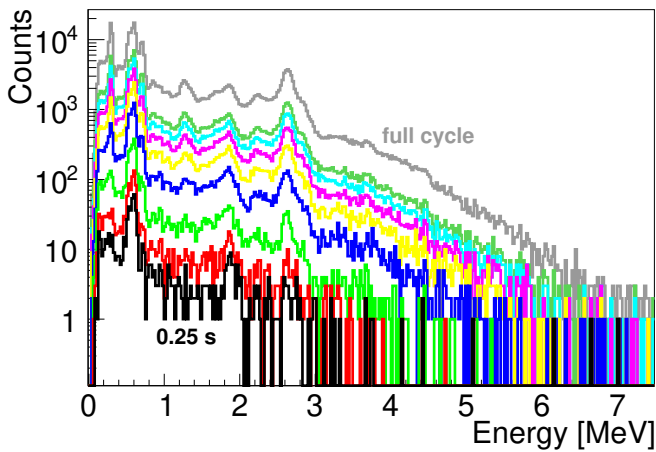}
\caption{Comparison of the DTAS $\beta$-gated spectra for the measurement of $^{102}$Zr+$^{102\text{m}}$Nb sorted with different time windows set within the total length of the cycle (11.6~s). The lengths of the time windows presented are: 0.25~s, 0.5~s, 1~s, 2~s, 3~s, 4~s, 5~s, 6~s and full cycle.}
\label{Cycles} 
\end{center}
\end{figure}

For each time window we calculate the contribution of each decay to the total TAGS spectrum by solving the corresponding Bateman equations. With two time windows, after subtracting the summing-pileup, we can calculate the individual contributions to the combined spectra by solving linear equations for the content of each bin in the TAGS spectrum. We have applied several time windows, as shown in Fig.~\ref{Cycles} for $^{102}$Zr+$^{102\text{m}}$Nb, and we have carried out the decomposition for every pair of spectra. Then we have computed the average of all the extracted spectra (weighted by a factor dependent on the statistics of the original spectra). 

\begin{figure}[h]
\begin{center}
\includegraphics[width=0.5 \textwidth]{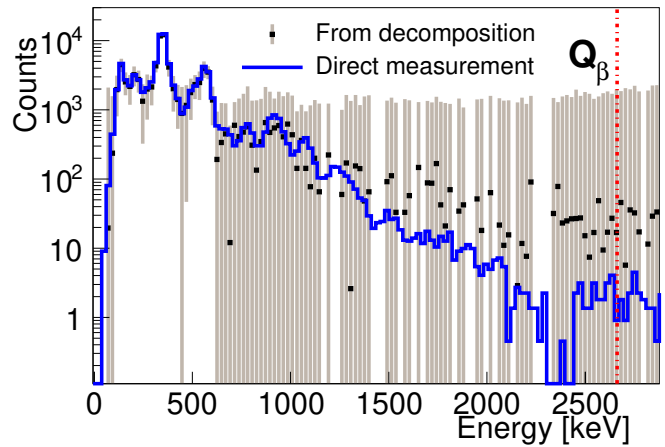}
\caption{Spectrum of the decay of $^{103}$Tc extracted with different time windows from the measurement of the decay of $^{103}$Mo (black dots with errors). The spectrum is compared with a direct measurement of the decay of $^{103}$Tc (blue).}
\label{103} 
\end{center}
\end{figure}

As a proof of the validity of this method, we have applied the same procedure to the measurement of the decay of $^{103}$Mo ($T_{1/2}$=67.5~s), where the activity of the daughter, $^{103}$Tc ($T_{1/2}$=54.2~s), is a contaminant. Both decays were measured independently in the same experimental campaign as the present niobium cases, with analogous good separation in JYFLTRAP. Preliminary results of their TAGS analyses were presented in~\cite{Acta103}. We have applied the present decomposition method to extract the spectrum of the decay of $^{103}$Tc from the measurement of the decay of $^{103}$Mo. In Fig.~\ref{103} the resulting spectrum is compared with the experimental spectrum of $^{103}$Tc measured independently. The agreement is excellent up to 600~keV, although from this energy onward the statistical fluctuations are dominant, showing the main limitation of this method. The errors associated with the Poisson statistics of the total spectra have been propagated to obtain the final error of each bin. Since statistics is the dominant source of error, we neglect the uncertainty of the weights calculated with the Bateman equations, as well as the uncertainty associated with the precision of the time window.

Using this procedure, we were able to separate the contributions of the decays of $^{100}$Zr and $^{102}$Zr to the spectra in each of the Zr+Nb combined measurements. The extracted spectra of the Zr decays exhibited large statistical fluctuations. For this reason, in the analyses of the low-spin niobium isomers, instead of using these extracted spectra as contaminants, we preferred to use smooth MC simulations that reproduce them. As will be explained below, it was possible to obtain information about the $\beta$ intensities of the Zr decays from the analysis of the extracted spectra that allowed us to perform such simulations. 

\begin{figure}[h]
\begin{center}
\includegraphics[width=0.5 \textwidth]{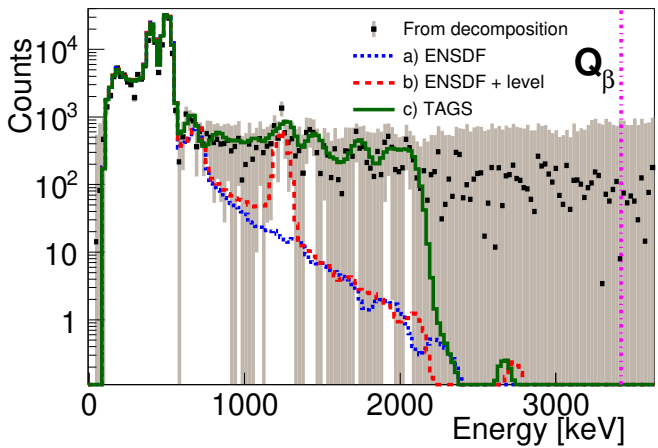}
\caption{Spectrum of the decay of $^{100}$Zr extracted with different time windows (dots with error bars) as explained in the text. The spectrum is compared with MC simulations of the decay of $^{100}$Zr assuming different $\beta$-intensity distributions as input for the DECAYGEN event generator \cite{TAS_decaygen} (see text).}
\label{100Zr_TAGS} 
\end{center}
\end{figure}

In Fig.~\ref{100Zr_TAGS} the extracted spectrum for the decay of $^{100}$Zr is shown. We have compared it with a MC simulation that uses the DECAYGEN event generator \cite{TAS_decaygen} with the $\beta$-intensities available at ENSDF~\cite{NDS_A100} as input, coming from the previously mentioned high-resolution experiment \cite{Zr_Sami}. As can be seen in Fig.~\ref{100Zr_TAGS} (dotted blue line), the known $\beta$-intensities reproduce the spectrum up to the last level known (704.1~keV), but there is a clear indication of a small decay contribution to states at higher energy. A study of the $^{100}$Mo(t,$^3$He)$^{100}$Nb reaction \cite{100Nb_more_levels} identified levels in $^{100}$Nb up to a level at 1300(30)~keV. We found that including in the simulation a $\beta$-intensity of 1$\%$ to a level at 1240~keV reproduces the peak observed in the extracted spectrum at this energy, as can be seen in Fig.~\ref{100Zr_TAGS} (dashed red line). Finally, to get an upper limit for the missing $\beta$-intensity, we also performed a TAGS analysis of the extracted spectrum. The level scheme of $^{100}$Nb used for the analysis included all those levels from ENSDF \cite{NDS_A100} identified up to 703.6~keV, ignoring the $5^+$ isomer. The $1^+$ assignment to the level at 471.38~keV was taken from RIPL-3 \cite{RIPL-3}, while for levels at 498.1~keV, 653.9~keV and 703.6~keV we have assumed spin-parity values of $0^+$. From this last level up to the $Q_{\beta}$ of the decay, a continuum of 40~keV bins based on the statistical model was used. We have analysed the extracted spectrum up to 2~MeV. In this analysis we restricted the feeding in the continuum region below 1.3~MeV to bins associated with those levels observed in \cite{100Nb_more_levels}. Due to the limited sensitivity of this analysis, we fixed the ground state feeding intensity to the 45$\%$ value from \cite{Zr_Sami}. The quality of the reproduction of the extracted spectrum is shown in Fig.~\ref{100Zr_TAGS} (solid green line), where the DECAYGEN event generator has been used to generate the TAGS spectrum with the branching ratio matrix and the $\beta$-intensities from the TAGS analysis as input.

\begin{figure}[h]
\begin{center}
\includegraphics[width=0.5 \textwidth]{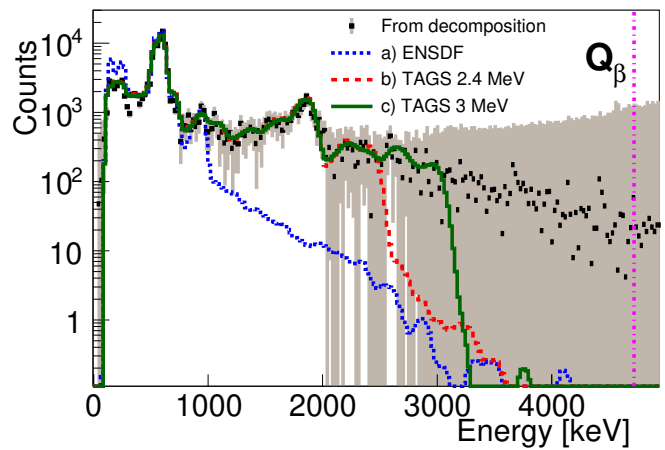}
\caption{Spectrum of the decay of $^{102}$Zr extracted with different time windows (dots with error bars) as explained in the text. The spectrum is compared with MC simulations of the decay of $^{102}$Zr assuming different $\beta$-intensity distributions as input for the DECAYGEN event generator \cite{TAS_decaygen} (see text).}
\label{102Zr_TAGS} 
\end{center}
\end{figure}

The extracted spectrum for the decay of $^{102}$Zr is shown in Fig.~\ref{102Zr_TAGS}. As in the previous case, there are some features not reproduced with a MC simulation that uses the results available at ENSDF \cite{NDS_A102} based on \cite{Zr_Sami} (dotted blue line in Fig.~\ref{102Zr_TAGS}). We have performed a TAGS analysis of the extracted spectrum, using as the known part of the level scheme of $^{102}$Nb the levels identified in the $\beta$-decay study from \cite{Zr_Sami} up to the level at 940.5~keV. A continuum of 40~keV bins based on the statistical model was used from this energy up to $Q_{\beta}$ of the decay. A summary of our spin-parity assumptions is presented in Table \ref{Table_levels_102Nb}. We have analyzed the extracted spectrum up to 2.4~MeV (dashed red line in Fig.~\ref{102Zr_TAGS}) and up to 3~MeV (solid green line in Fig.~\ref{102Zr_TAGS}). As in the case of $^{100}$Zr, the ground state feeding intensity was also fixed to the 59$\%$ value obtained in \cite{Zr_Sami}.

\begin{table}[H]
\centering
  \begin{tabular}{c|c|c}
Energy [keV] & $J^{\pi}$ ENSDF & $J^{\pi}$ used  \\ \hline
0.00 +x& $1^{+}$ & $1^+$ \\
20.37 +x& - & $1^+$ \\
64.39 +x& ($2^{+}$) & $2^+$ \\
93.95 +x& - & $2^+$ \\
156.36 +x& - & $2^+$ \\
160.72 +x& - & $2^+$ \\
246.31 +x& - & $1^-$ \\
258.43 +x& - & $1^-$ \\
430.70 +x& - & $1^-$ \\
599.49 +x& $1^{+}$ & $1^+$ \\
705.08 +x& (1) & $1^+$ \\
940.50 +x& (1) & $1^+$ \\
\hline
  \end{tabular}
\caption{Levels in $^{102}$Nb with no spin-parity assignment in ENSDF and the adopted values (up to 940.5~keV). The energy offset x is 94~keV according to the NUBASE evaluation of 2016 \cite{NUBASE2016} and 93(23)~keV from the experimental work of S. Rinta-Antila \textit{et al.}~\cite{Zr_Sami}.}
\label{Table_levels_102Nb}
\end{table}

\subsection{Shape studies of $^{100,102}$Zr}

Zirconium isotopes are known to have a phase transition around $^{100}$Zr from spherical to deformed ground states \cite{Federman_deformation}. Recently a quantum phase transition due to type II shell evolution has been predicted in this region from Monte Carlo shell model (MCSM) calculations \cite{ShapeZr2016}. Shape coexistence of spherical and deformed structures in $^{96}$Zr has been deduced \cite{Shape96Zr}, based on an electron scattering measurement. For $^{98}$Zr, mixing of spherical and deformed configurations has been suggested in a very recent work at Argonne National Laboratory (ANL)~\cite{Shape98Zr}.

The TAGS technique has proven to be useful to study shape effects in the parent nucleus, by comparing the experimental $B$(GT) strength distributions with theoretical calculations~\cite{KikeShape,PRC_Poirier,PRC_AnaBelen,PRC_Esther,PRC_Briz}. In order to provide experimental information to be compared with theoretical calculations, we have evaluated the $\beta$-strength distributions as a function of the excitation energy ($E_x$) for $^{100,102}$Zr. For this we have used Equation \eqref{srength_exp}:

\begin{equation}\label{srength_exp}
S_{\beta}(E_x)=\frac{I_{\beta}(E_x)}{f(Q_{\beta}-E_x,Z)T_{1/2}}
\end{equation}

\noindent where $f$ is the Fermi statistical rate function, evaluated employing subroutines from the $\log ft$ program of the National Nuclear Data Center (NNDC) \cite{logftNNDC}, $I_{\beta}$ is the normalized $\beta$ intensity determined from our experiment, $Q_{\beta}$ is the $\beta$ decay Q value and $T_{1/2}$ is the decay half-life. 

The theoretical calculations employed in the comparison are taken from the work of P. Sarriguren \textit{et al.} \cite{SarrigurenMoZr, SarrigurenMoZr_2010} and were performed in the A =100 - 120 region for neutron-rich molybdenum and zirconium isotopes. These calculations are performed within the proton-neutron quasiparticle random-phase approximation. This microscopic approach is based on a deformed Hartree-Fock + BCS mean field obtained with Skyrme interactions to generate single-particle energies, wave functions, and occupation probabilities. In particular, the SLy4 Skryme force is used. Spin-isospin residual interactions are included on top of this mean field.

\begin{figure}[h]
\begin{center}
\includegraphics[width=0.5 \textwidth]{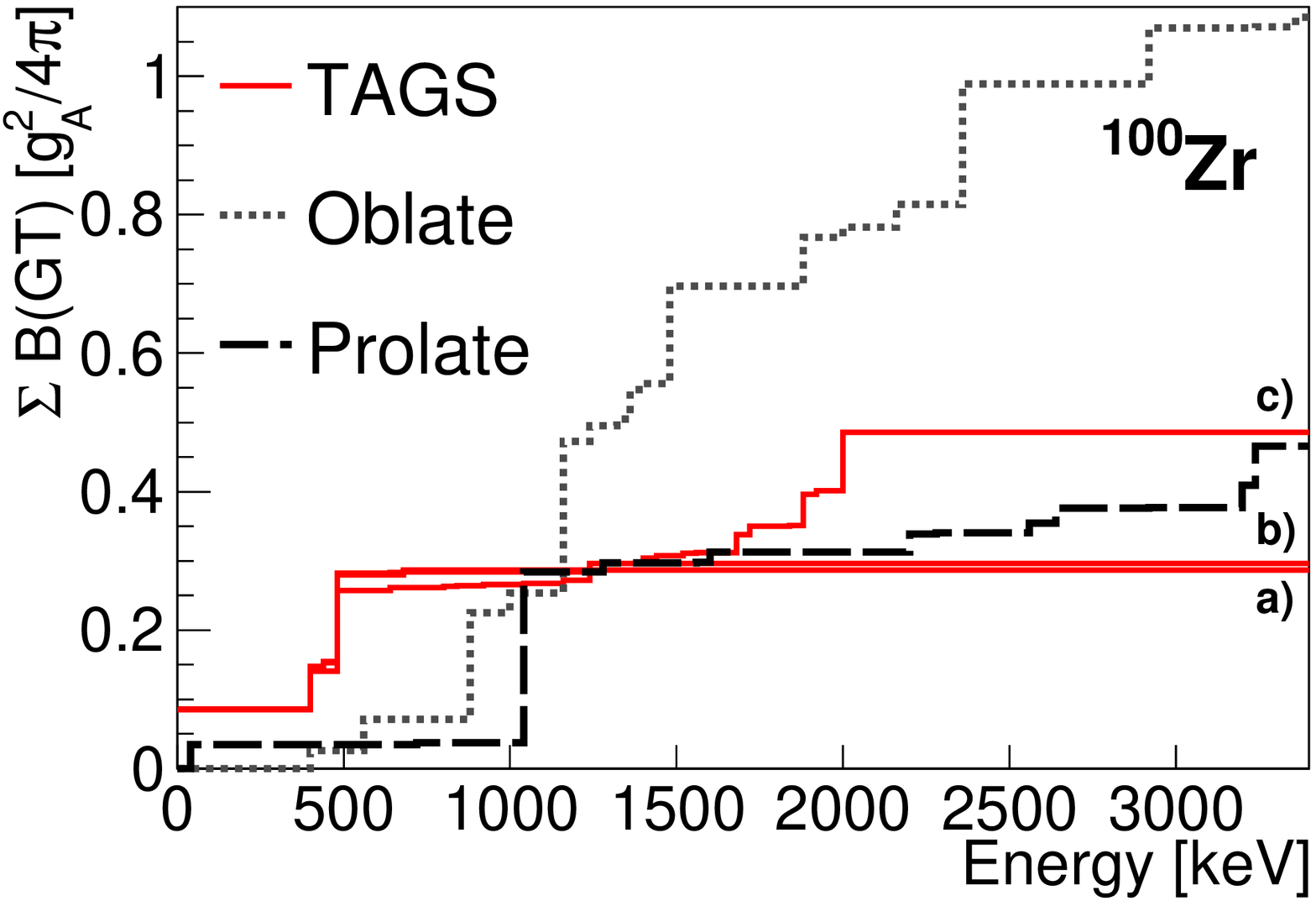}

\includegraphics[width=0.5 \textwidth]{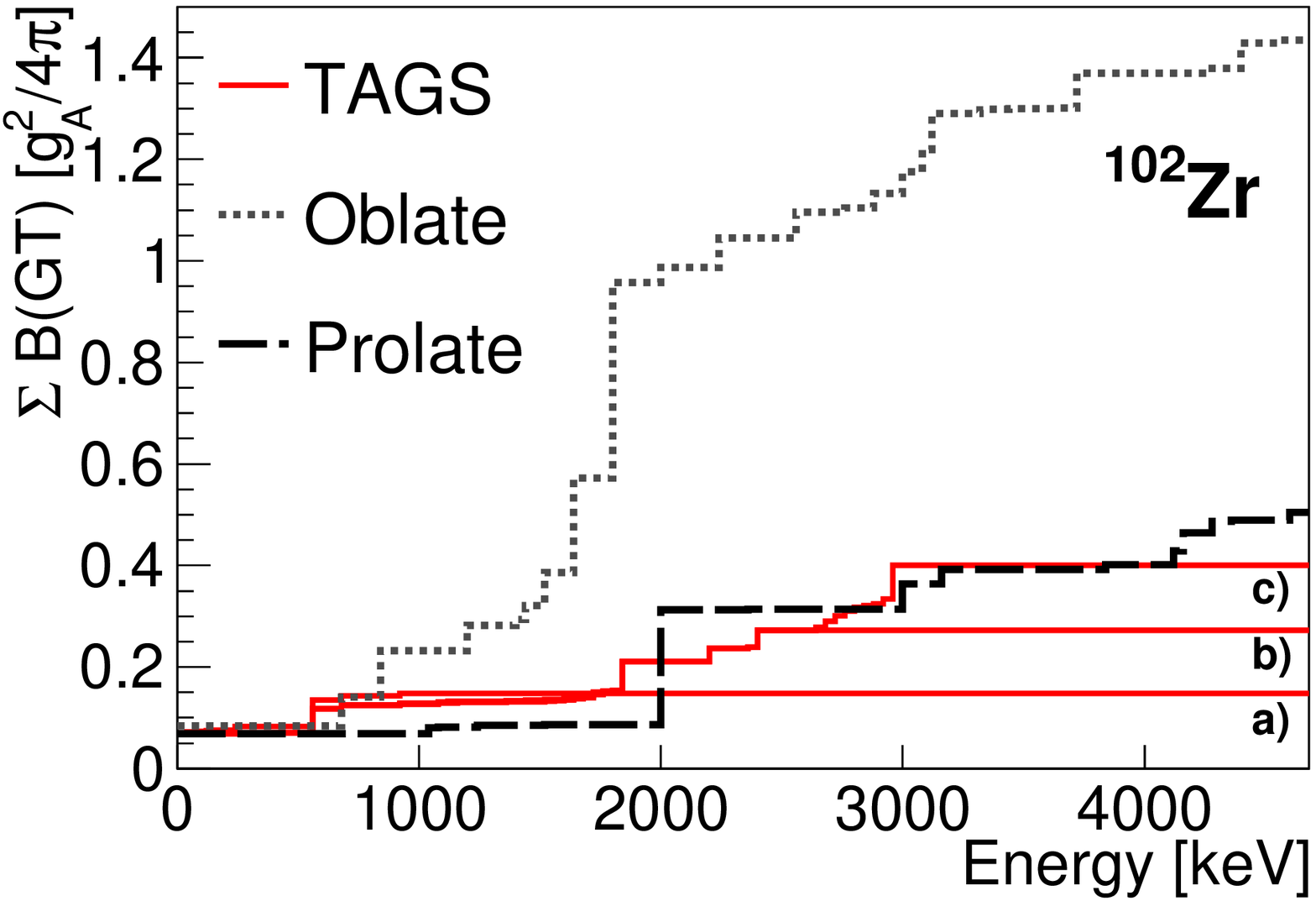}
\caption{Comparison of the experimental accumulated $B$(GT) distribution (solid red) with the theoretical distributions for oblate (dotted grey) and prolate (dashed black) configurations. Results for $^{100}$Zr (top) and $^{102}$Zr (bottom) are included. The experimental $B$(GT) distributions for three possible solutions are considered for each case (see text and Figs.~\ref{100Zr_TAGS} and \ref{102Zr_TAGS}).}
\label{BGT} 
\end{center}
\end{figure}

Experimental $B$(GT) distributions are compared with the theoretical calculations in Fig.~\ref{BGT}. The calculations have been scaled by a quenching factor for the comparison ($(g_A/g_V)_{\text{eff}}=0.77(g_A/g_V)_{\text{free}}$). We have evaluated the $B$(GT) distribution for the three possible solutions considered for each case in Figs. \ref{100Zr_TAGS} and \ref{102Zr_TAGS}, as a sort of upper and lower limit. Even though our TAGS analyses of the zirconium isotopes represent just an estimate, the comparison with the very different prolate and oblate strength patterns obtained in the theoretical calculations \cite{SarrigurenMoZr,SarrigurenMoZr_2010}, suggests a clear  dominance of the prolate configuration in both cases. 

\section{Analyses of $^{100\text{gs},100\text{m}}$N\lowercase{b}\label{100Nb}}

As explained in Section \ref{exp}, the decay of the $^{100\text{gs}}$Nb was studied from the parent $^{100}$Zr implantation measurement. The decay of $^{100}$Zr was thus considered as a contamination using the information obtained in the previous section. In a subsequent step, the decay of $^{100\text{m}}$Nb was studied from a $^{100}$Nb implantation measurement that contains both isomers, where the $^{100\text{gs}}$Nb was treated as a contaminant.

In the calculation of the response functions for the analyses of the $^{100}$Nb isomers we used the known information from ENSDF at low excitation energies \cite{NDS_A100}. According to the recommendations of RIPL-3 \cite{RIPL-3}, the level scheme of $^{100}$Mo is complete up to the level at 2339.8~keV. From this level up to the $Q_{\beta}$ value, a continuum region with 40~keV bins was used, as explained before. For those levels without assigned spin-parity values in the known region, the recommended values from RIPL-3 \cite{RIPL-3} were employed.

\subsection{$^{100\text{gs}}$Nb}

For the analysis of the decay of the low-spin ground state (1$^+$), we considered allowed transitions plus forbidden transitions to levels at 1136~keV ($4^+$), 1607.36~keV ($3^+$), 1908.28~keV ($3^{-}$) and 2189.53~keV ($4^{+}$) in $^{100}$Mo (recommended spin-parity values in parentheses). Feeding to all these levels was seen in high resolution measurements, with the exception of the level at 1908.28~keV, whose inclusion was found to improve the analysis. Except for the level at 1136~keV, with a firm spin-parity assignment, the spin-parity of the three remaining levels is compatible with a $3^{-}$ assignment (first forbidden transitions)~\cite{NDS_A100}. The previously reported $\beta$-intensity to the level at 1136~keV~\cite{NDS_A100} is questionable due to the proximity of a peak in the decay of $^{100}$Zr. For this reason, we studied the impact of not allowing feeding to this level, and a similar result was obtained. Finally, feeding to the level at 2189.53~keV is indistinguishable from feeding to the level at 2201.12~keV ($2^-$), not seen in high resolution measurements. 

In the decay of the $^{100\text{gs}}$Nb the decay of $^{100}$Zr was treated as a contamination, as mentioned above. In order to avoid the effect of the large statistical fluctuations of the extracted spectrum presented in Section \ref{Zr}, this contribution was calculated with MC simulations, by using the DECAYGEN event generator \cite{TAS_decaygen} with the $\beta$-intensities discussed in the previous section. Given the large uncertainty at high energies in the decomposition method used for the $^{100}$Zr contaminant, we opted to take the solution b) in Fig.~\ref{100Zr_TAGS} as the reference shape for the $^{100}$Zr contaminant. This solution is intermediate to solutions a) and c) which are then used to estimate the systematic uncertainty associated with this contaminant as explained below. The normalization of this contamination was obtained using its most intense $\gamma$-ray, which connects the level at 400.5~keV with the ground state in $^{100}$Nb.

\begin{figure}[!hbt]
\begin{center}
\includegraphics[width=0.5 \textwidth]{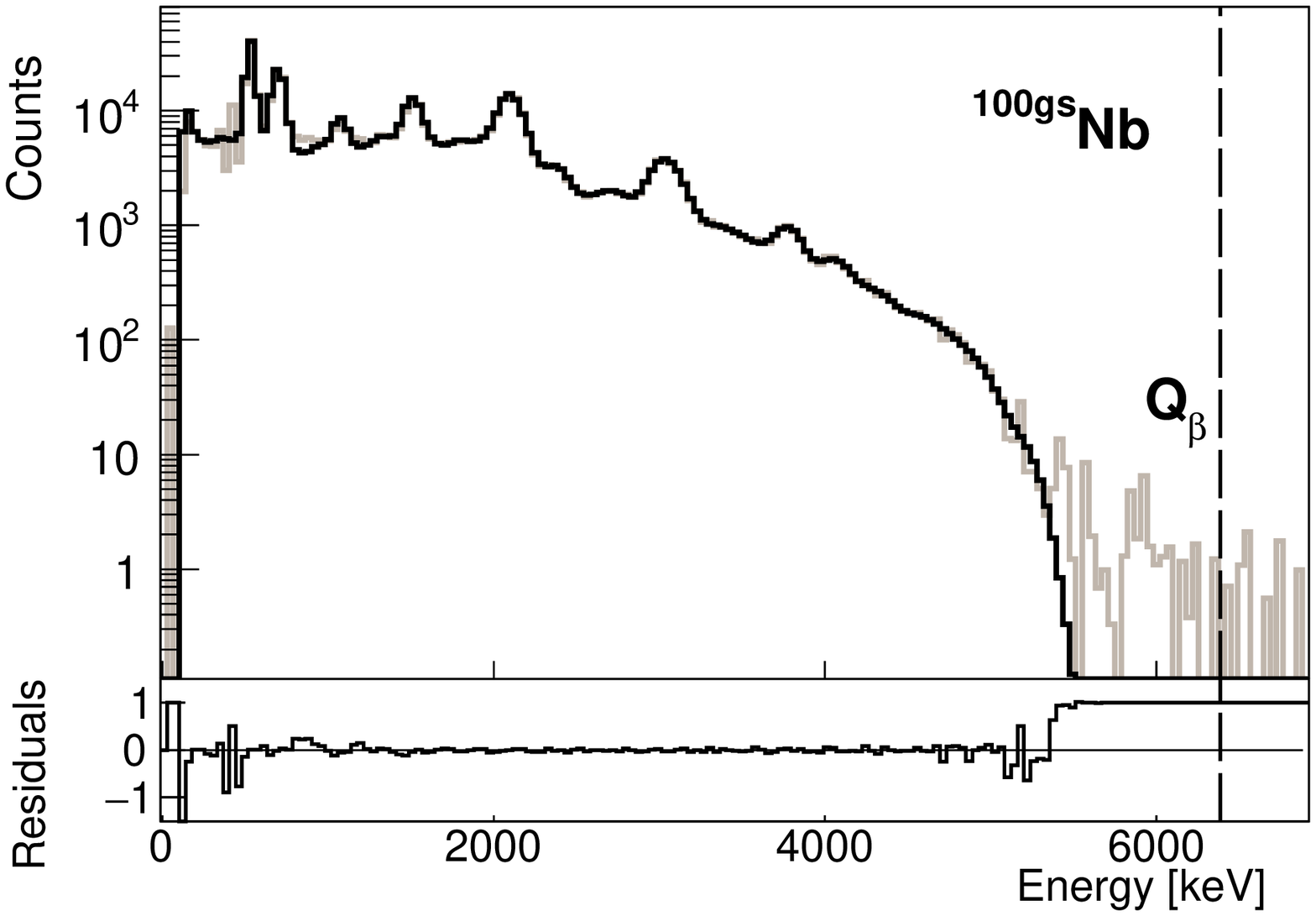} \\
\includegraphics[width=0.5 \textwidth]{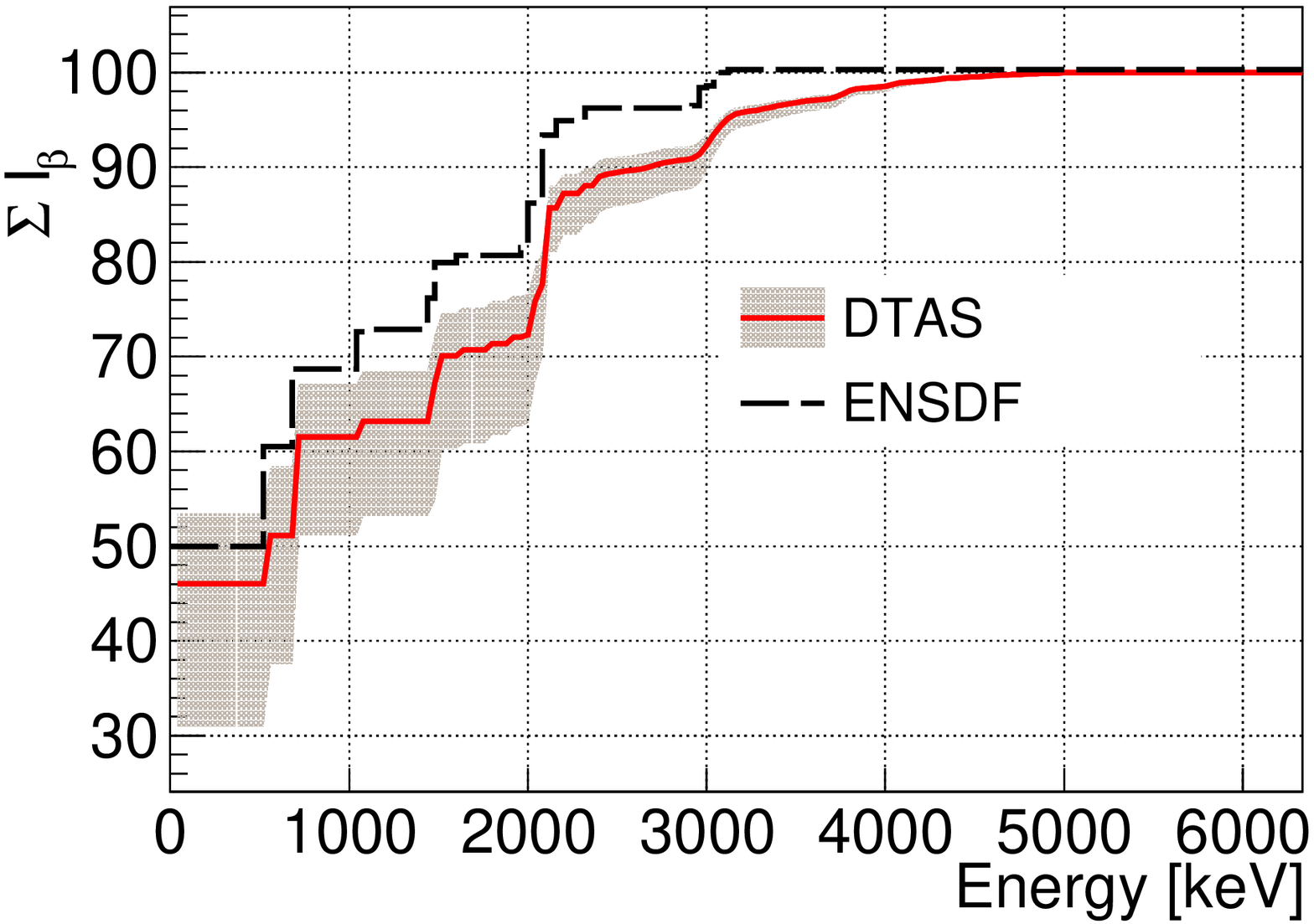} 
\caption{Top panel: experimental $\beta$-gated spectrum free of contaminants for $^{100\text{gs}}$Nb (grey) and reconstructed spectrum (black). The relative deviations between experimental and reconstructed spectra are also shown. Bottom panel: accumulated $\beta$-intensity for the present TAGS results (solid line) and for the data from ENSDF (dashed line). The filled region represents the systematic uncertainty (see text for details).} 
\label{100gsNb_fit}
\end{center}
\end{figure}

In the top panel of Fig.~\ref{100gsNb_fit}, the quality of the TAGS analysis can be seen by comparing the experimental spectrum free of contaminants with the spectrum reconstructed by applying Equation \eqref{inverse}, i.e. the $\beta$ intensities obtained in the analysis convoluted with the corresponding response function of the spectrometer. The accumulated $\beta$-intensity obtained from the analysis is shown in Fig.~\ref{100gsNb_fit} bottom panel. A slight Pandemonium effect can be deduced from this figure when comparing the TAGS results with the accumulated $\beta$-intensity from ENSDF~\cite{NDS_A100}. The ground state (g.s.) feeding intensity obtained from the TAGS analysis, $46_{-15}^{+8}\%$, is compatible within the errors with the value from ENSDF, 50(7)$\%$. We have also applied a $\beta$-$\gamma$ counting method for TAGS data proposed by Greenwood et al. \cite{Greenwood_GS}, which gives a result of 41(16)$\%$, also pointing to a $\beta$-intensity to the ground state lower than the value quoted by ENSDF, but compatible with it as well. The $\beta$ intensity distribution of the Nb cases studied in this work can be found in the Supplemental Material~\cite{Suplement}.

In all the cases presented here the uncertainties that we are quoting are of systematic character, since statistical ones are negligible in comparison. Some sources of systematic error have been considered for all the analyses presented in this work (and will not be mentioned again for the following cases): errors in the energy and resolution calibrations, the effect of the threshold of the $\beta$-detector (which affects the energy dependence of the efficiency of this detector) and the possible effect of the deconvolution algorithm. To evaluate the latter, we employed the Maximum-Entropy algorithm in addition to the Expectation-Maximization one, usually employed in our analyses~\cite{TAS_algorithms}. Additionally, for the decay of $^{100\text{gs}}$Nb, all the possibilities mentioned above for the branching ratio matrix were considered for the estimation of the uncertainties, including a branching ratio matrix with a $3^{-}$ assignment for the levels at 1607.36~keV, 1908.28~keV and 2189.53~keV. The three possible spectral shapes discussed in Section \ref{Zr} for the contamination of the decay of $^{100}$Zr were used. For each of them, a change in the normalization factor of around 10$\%$ was shown to be compatible with a reasonable reproduction of the spectrum, and it affects the g.s. feeding obtained. The reproduction of the spectrum was also compatible with a change in the normalization of the summing-pileup of up to 20$\%$. The effect of the first bin included in the analysis was also investigated, showing a large impact in the g.s. feeding intensity, comparable to the impact of the normalization of the parent activity. Finally, a solution obtained with a branching ratio matrix modified to reproduce the known $\gamma$-intensities for low-lying levels was also taken into consideration. In Table \ref{tableIg100Nb}, the initial $\gamma$-intensities and the ones obtained with the modified branching ratio matrix are presented. The original $\gamma$-intensities from the TAGS analysis are not far from the high resolution values \cite{NDS_A100}, also included in Table \ref{tableIg100Nb}, and the reproduction of the experimental TAGS spectrum barely changes when this modified branching ratio matrix is used.

\begin{table*}[h]
\centering
\begin{tabular}{|c|ccc|ccc|}
 & \multicolumn{3}{c|}{$^{100\text{gs}}$Nb} & \multicolumn{3}{c|}{$^{100\text{m}}$Nb} \\
Energy [keV] & $I_{\gamma}$ ENSDF & $I_{\gamma}$ DTAS & $I_{\gamma}$ DTAS$^{*}$ & $I_{\gamma}$ ENSDF & $I_{\gamma}$ DTAS & $I_{\gamma}$ DTAS$^{*}$ \\ \hline
536	&	0.457	& 0.463	& 0.457   &  0.950 &  0.910 & 0.960  \\
695  &	0.088	& 0.129 & 0.123   &  0.036 &  0.025 & 0.021  \\
1064  &	0.125 	& 0.133 & 0.125   &  0.141 &  0.289 & 0.136  \\
1136  &	0.006 	& 0.005 & 0.007   &  0.694 &  0.358 & 0.696  \\
1464  &	0.062 	& 0.073 & 0.077   &  0.085 &  0.046 & 0.043  \\
1505  &	0.037 	& 0.031 & 0.032   &     -  &    -    & -     \\ 
1607  &	0.010 	& 0.010 & 0.010   &  0.098 &  0.118 & 0.070  \\
1771  &     	-	& -	& -	  &  0.096 &  0.043 & 0.023  \\
1847  &	 -	& -	& -	  &  0.065 &  0.039 & 0.024  \\
1977  &	0.008 	& 0.003 & 0.004   &     -  &    -    & -     \\
2038  &	0.047 	& 0.036 & 0.050   &     -  &    -    & -     \\
2086  &	0.072 	& 0.081 & 0.082   &     -  &    -    & -     \\
2103  &	-	& -     & -       &  0.263 &  0.155 & 0.257   \\
2190   &	0.015 	& 0.015 & 0.016   &     -  &    -    & -     \\	
2289   &	-	& -     & -       &  0.018 &  0.061 & 0.037   \\
2310  &	-	& -     & -       &  0.093 &  0.057 & 0.038  \\
2320  &	0.013 	& 0.001 & 0.001   &     -  &    -    & -     \\					
\hline
\end{tabular}
\caption{\label{tableIg100Nb} Absolute $\gamma$-intensities (per 100 decays) de-exciting the main levels populated in the decays of $^{100\text{gs},100\text{m}}$Nb to $^{100}$Mo. The ENSDF column corresponds to the intensities obtained from high resolution data \cite{NDS_A100}. The intensities obtained with the original branching ratio matrix (DTAS) and those obtained with a modified branching ratio matrix (DTAS$^{*}$) are presented for each case.}
\end{table*}

\subsection{$^{100\text{m}}$Nb}

\begin{figure}[h]
\begin{center}
\includegraphics[width=0.5 \textwidth]{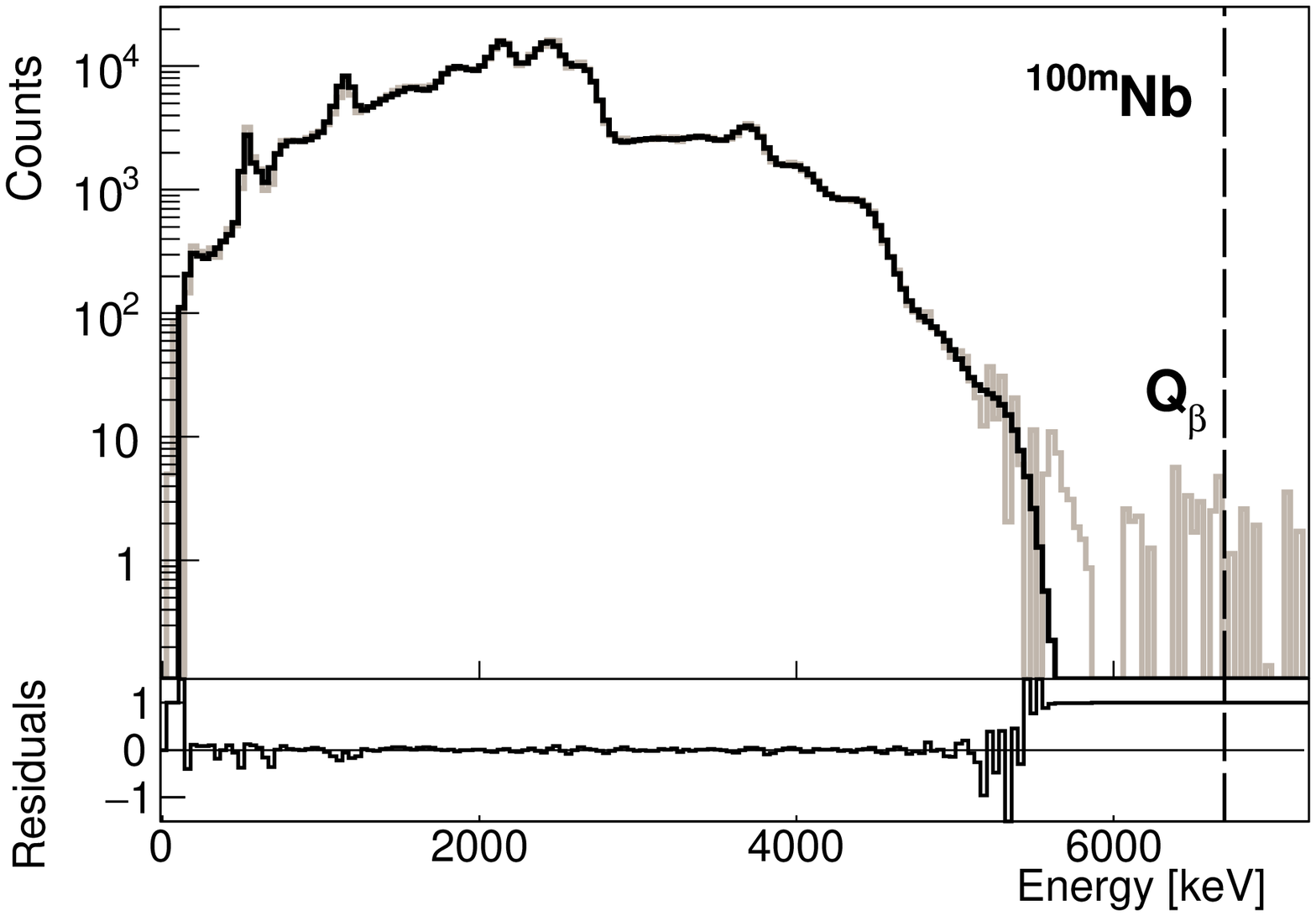} \\
\includegraphics[width=0.5 \textwidth]{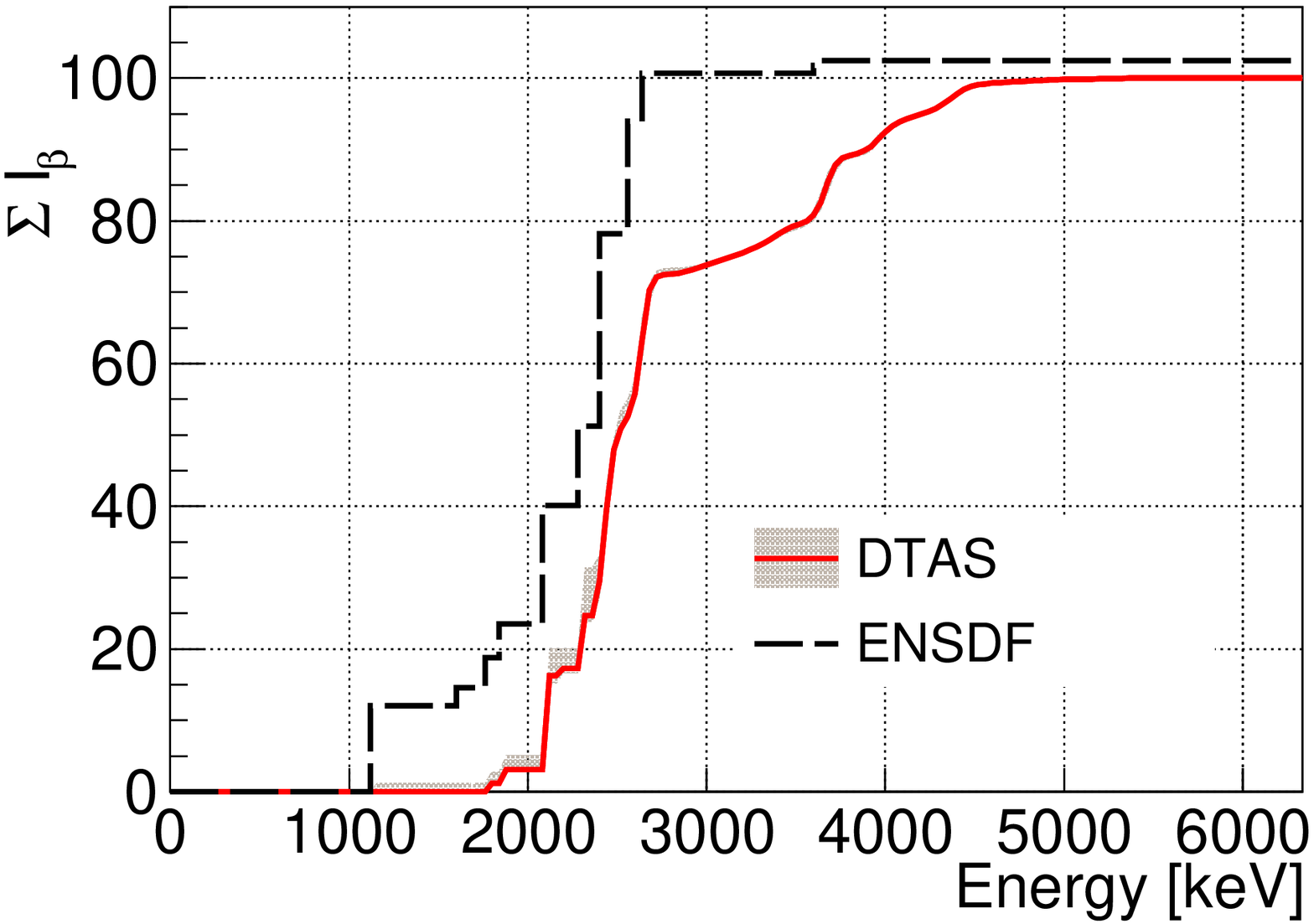} 
\caption{Same caption as in Fig.~\ref{100gsNb_fit} but for $^{100\text{m}}$Nb.} 
\label{100mNb_fit}
\end{center}
\end{figure}

For the calculation of the response function of the decay of the high-spin (5$^{+}$) isomer we took into account allowed transitions to 4$^{+}$, 5$^{+}$ and 6$^{+}$ states in $^{100}$Mo. The contribution of the decay of the low-spin ground state was treated as a contaminant in the $^{100}$Nb implantation measurement that contains both isomers. This contribution was normalized with the peak at 695.2~keV, associated with the de-excitation of a 0$^+$ level populated only in the decay of the low spin isomer. A check of the normalization is provided by the shape of the spectrum at low energies. This is affected by the penetration in the detector of g.s. to g.s $\beta$ particles in the decay of the $^{100\text{gs}}$Nb, but not in the case of the high-spin isomer (compare Fig.\ref{100gsNb_fit} top and Fig. \ref{100mNb_fit} top). 

In this case the reproduction of the experimental TAGS spectrum, shown in Fig.~\ref{100mNb_fit} top, is significantly improved when a modified branching ratio matrix is considered. This modified branching ratio matrix is calculated to reproduce the $\gamma$-intensities for low energy levels coming from high resolution data \cite{Suhonen_100mNb,100Nb100mNb_1987}, and presented in Table \ref{tableIg100Nb}. The $\gamma$-intensities obtained from the TAGS analysis before and after modifying the branching ratio matrix are also listed. We have used this modified branching ratio matrix for the determination of the reference $\beta$-intensity distribution of the TAGS analysis. 

The systematic uncertainties due to changes in the normalization factors of the contaminants has been estimated, with a change of 10$\%$ in both the summing-pileup and in the contamination of the $^{100\text{gs}}$Nb. The original branching ratio matrix and the modified one were also taken into account in the evaluation of the errors. We have also considered the effect of subtracting the experimental low-spin spectrum (free of contaminants) or a simulated one, generated with the results of the TAGS analysis. For the experimental one we have taken into account the resulting spectra obtained with the three spectral shapes for $^{100}$Zr discussed above (see Fig. \ref{100Zr_TAGS}). All these contributions to the error budget give the uncertainty in the accumulated $\beta$-intensity shown in Fig.~\ref{100mNb_fit} bottom, where a clear Pandemonium effect is seen when comparing it with ENSDF data.

\begin{figure}[h]
\begin{center}
\includegraphics[width=0.5 \textwidth]{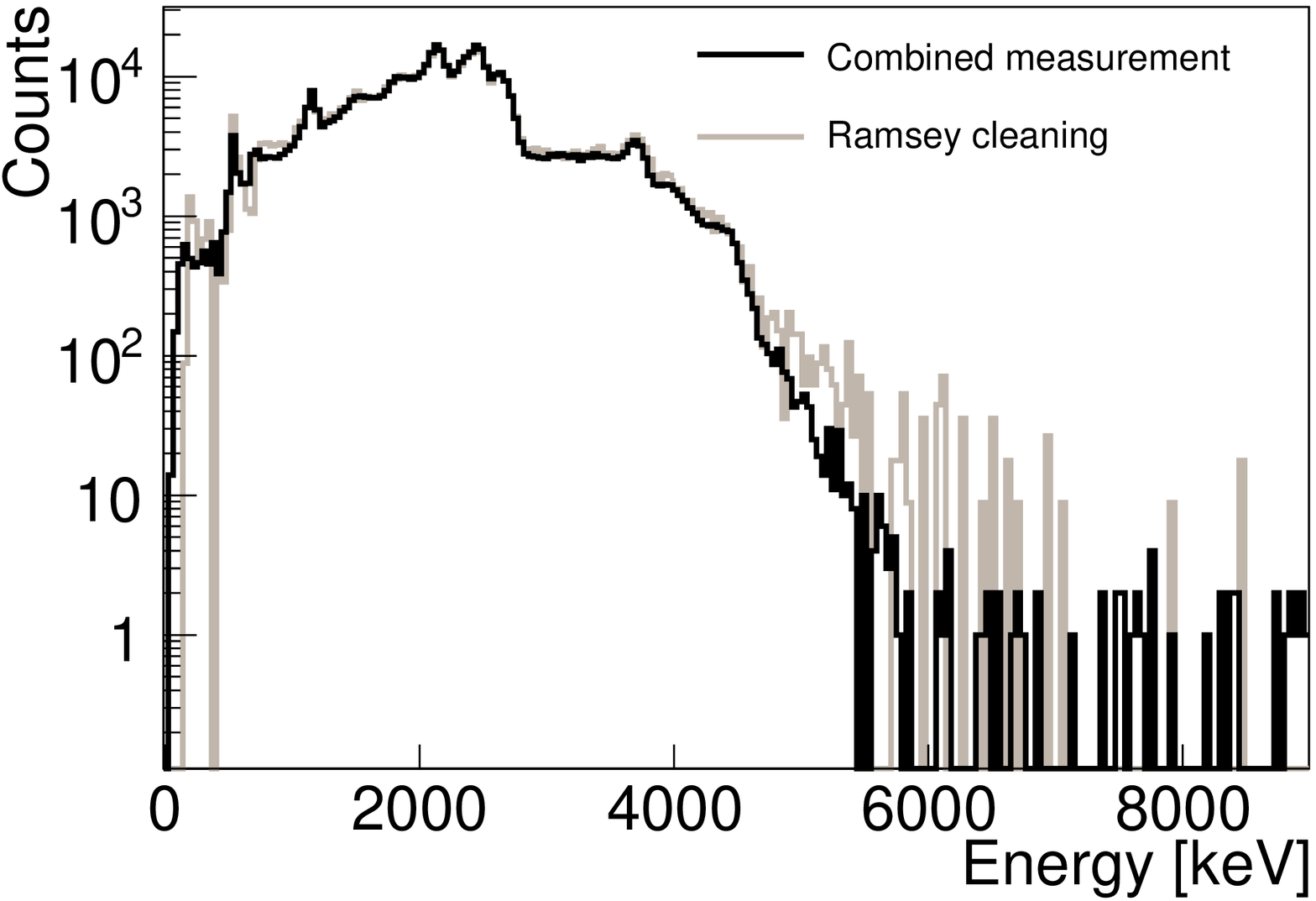} \\

\includegraphics[width=0.5 \textwidth]{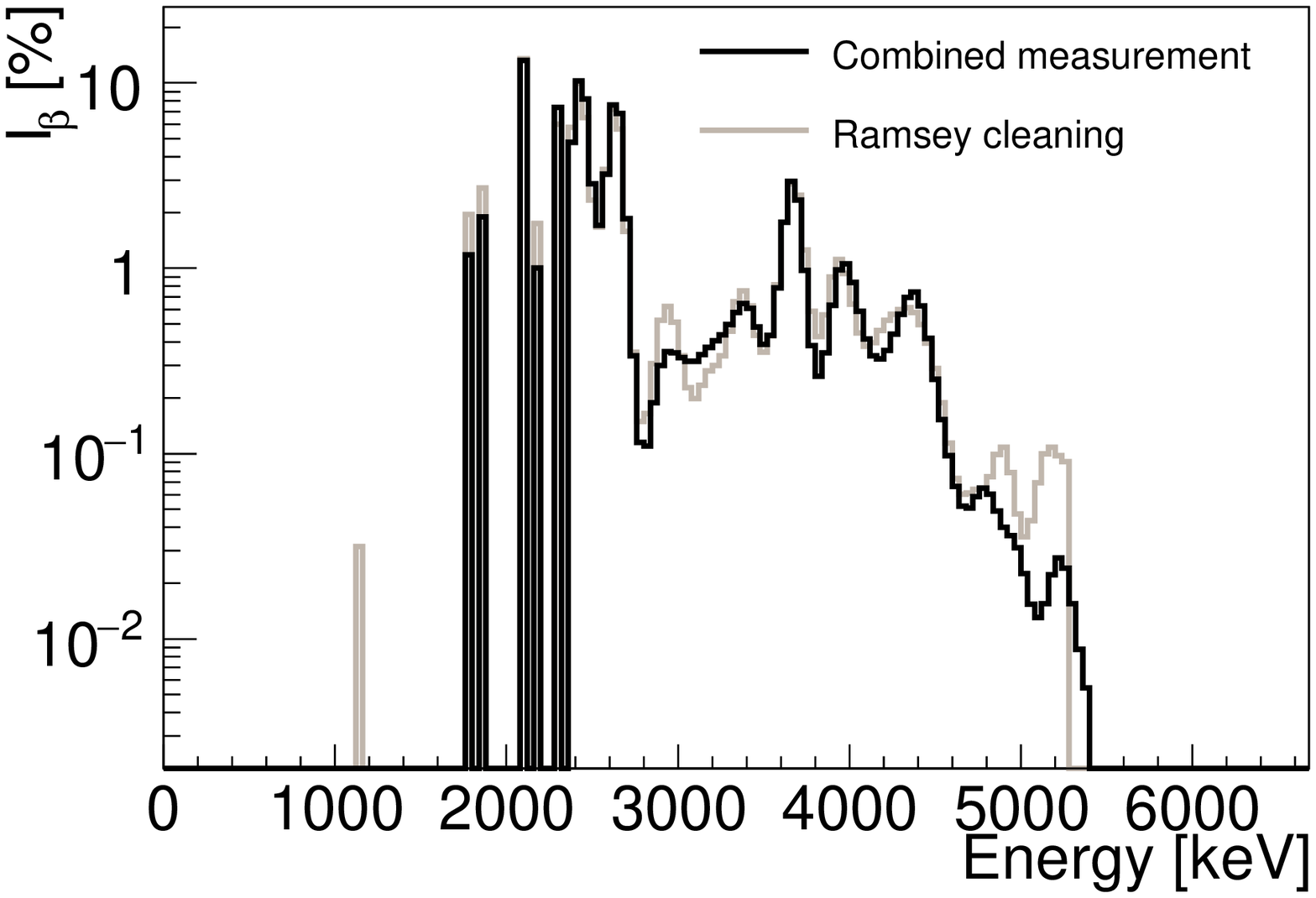} 
\caption{Top: experimental spectra for the decay of $^{100\text{m}}$Nb free of contaminants. The back line corresponds to the spectrum coming from the combined measurement of the two isomers. The gray line corresponds to the spectrum obtained after applying the Ramsey cleaning purification.  Bottom: comparison of the $\beta$ intensity distributions obtained from both experimental spectra.} 
\label{100mNb_combined_vs_Ramsey}
\end{center}
\end{figure}

As a cross-check of the separation of the isomers, we have employed the high resolution purification Ramsey cleaning technique in JYFLTRAP~\cite{Ramsey}. It is based on the use of the purification trap for a first isobaric cleaning, followed by the use of the precision trap for isomeric cleaning. A run with limited statistics was measured setting the frequency to select the high-spin isomer. The study of the resulting spectrum free of summing-pileup gave a contamination of 16.5$\%$ from the $^{100\text{gs}}$Nb. A 9.3$\%$ contribution from the decay of $^{100}$Zr was also found, due to an accidental overlap between the frequency selected in the purification process for $^{100\text{m}}$Nb and the repeating frequency corresponding to $^{100}$Zr, as reported in \cite{100Nb100mNb_IGISOL}. The contributions of the low-spin $^{100\text{gs}}$Nb branch and of the decay of $^{100}$Zr were normalized as mentioned before. A small environmental background contamination was also taken into account because of random coincidences due to the low counting rate. A comparison of the spectrum after subtracting all these contaminants with the one corresponding to the combined measurement of the two isomers (also free of contaminants) is shown in Fig.~\ref{100mNb_combined_vs_Ramsey} top. Note that the latter has been normalized to the number of counts of the first one -with a factor 9 less statistics- for comparison. The $\beta$-intensity distribution obtained from the analysis of the Ramsey-cleaned spectrum is compared with the one obtained from the analysis of the combined measurement in Fig.~\ref{100mNb_combined_vs_Ramsey} bottom. Both results are in reasonable agreement.

\section{Analyses of $^{102\text{gs},102\text{m}}$N\lowercase{b}\label{102Nb}}

Similarly to the A=100 case, the decay of $^{102\text{m}}$Nb was studied from the parent $^{102}$Zr implantation measurement, as explained in Section \ref{exp}. The decay of $^{102}$Zr was considered as a contaminant using the information obtained in Section \ref{Zr}. The decay of $^{102\text{gs}}$Nb was studied from the $^{102}$Nb implantation measurement that contains both isomers, where $^{102\text{m}}$Nb was considered as a contamination. 

In the TAGS analyses of the A=102 niobium cases, we have considered the known level scheme in $^{102}$Mo up to the level at 1398.39~keV excitation energy, according to the recommendations of RIPL-3 \cite{RIPL-3}. The uncertain spin-parity values were taken from the values proposed in RIPL-3 \cite{RIPL-3}.

\subsection{$^{102\text{m}}$Nb}

For the analysis of the decay of the low-spin isomer (1$^+$), we considered allowed transitions to positive parity levels with spins 0, 1 and 2. Direct feeding at around 1330~keV was required in order to improve the reproduction of the experimental TAGS spectrum at this excitation energy. Although there is a 0$^{+}$ level at 1334.5~keV, which would be easily fed in the decay, no $\gamma$-rays had ever been seen from this level, identified in a $^{100}$Mo(t,p) reaction. We have assumed that it decays via a two-$\gamma$ cascade. The spectral shape b) from Fig.~\ref{102Zr_TAGS} obtained for the decay of $^{102}$Zr was used as a contaminant in order to avoid the statistical fluctuations of the extracted $^{102}$Zr spectrum. The normalization of this contribution was obtained using the 599.48~keV and 535.13~keV $\gamma$-rays, emitted in the de-excitation of the 1$^{+}$ level at 599.48~keV in $^{102}$Nb. 

\begin{figure}[h]
\begin{center}
\includegraphics[width=0.5 \textwidth]{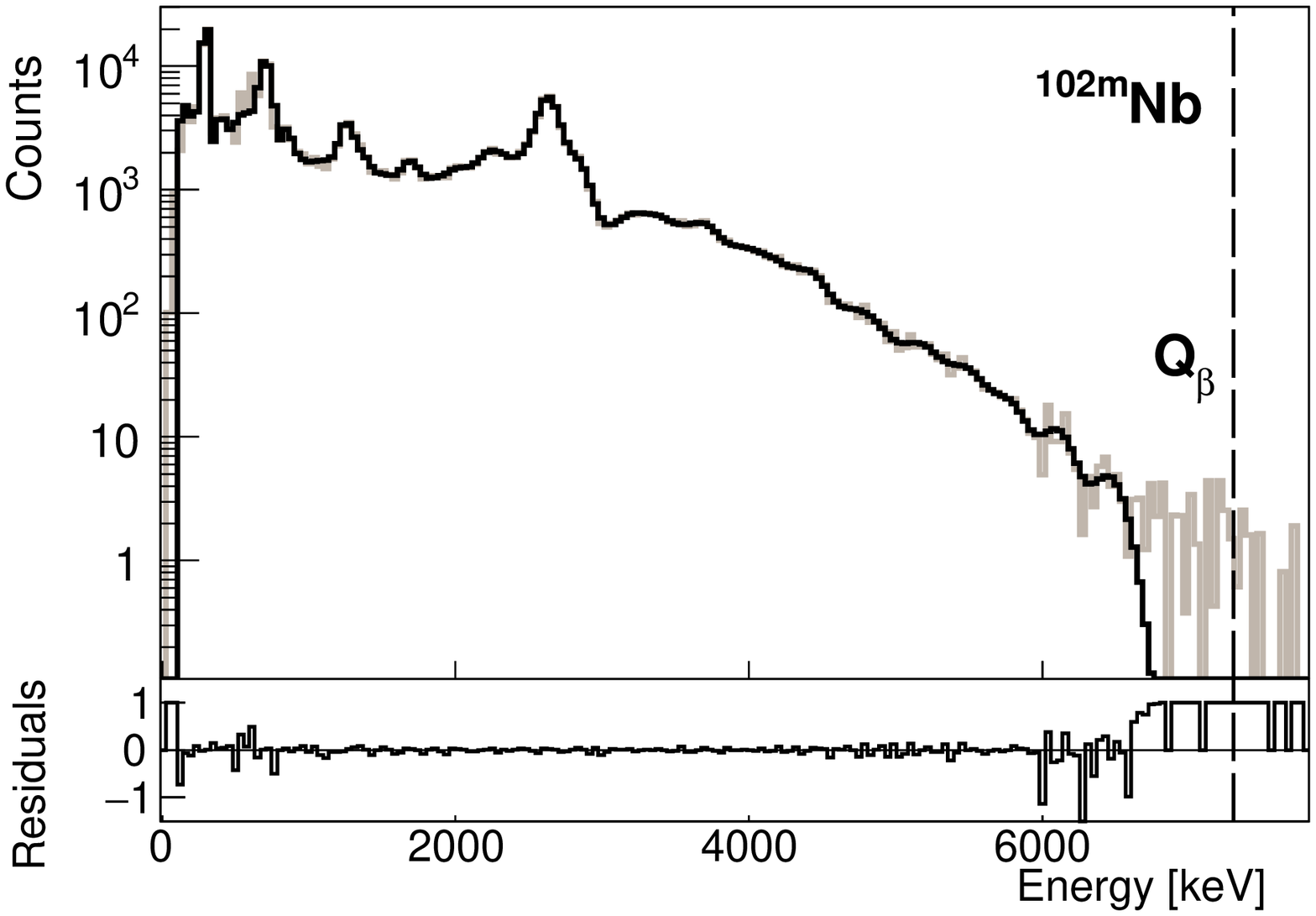} \\
\includegraphics[width=0.5 \textwidth]{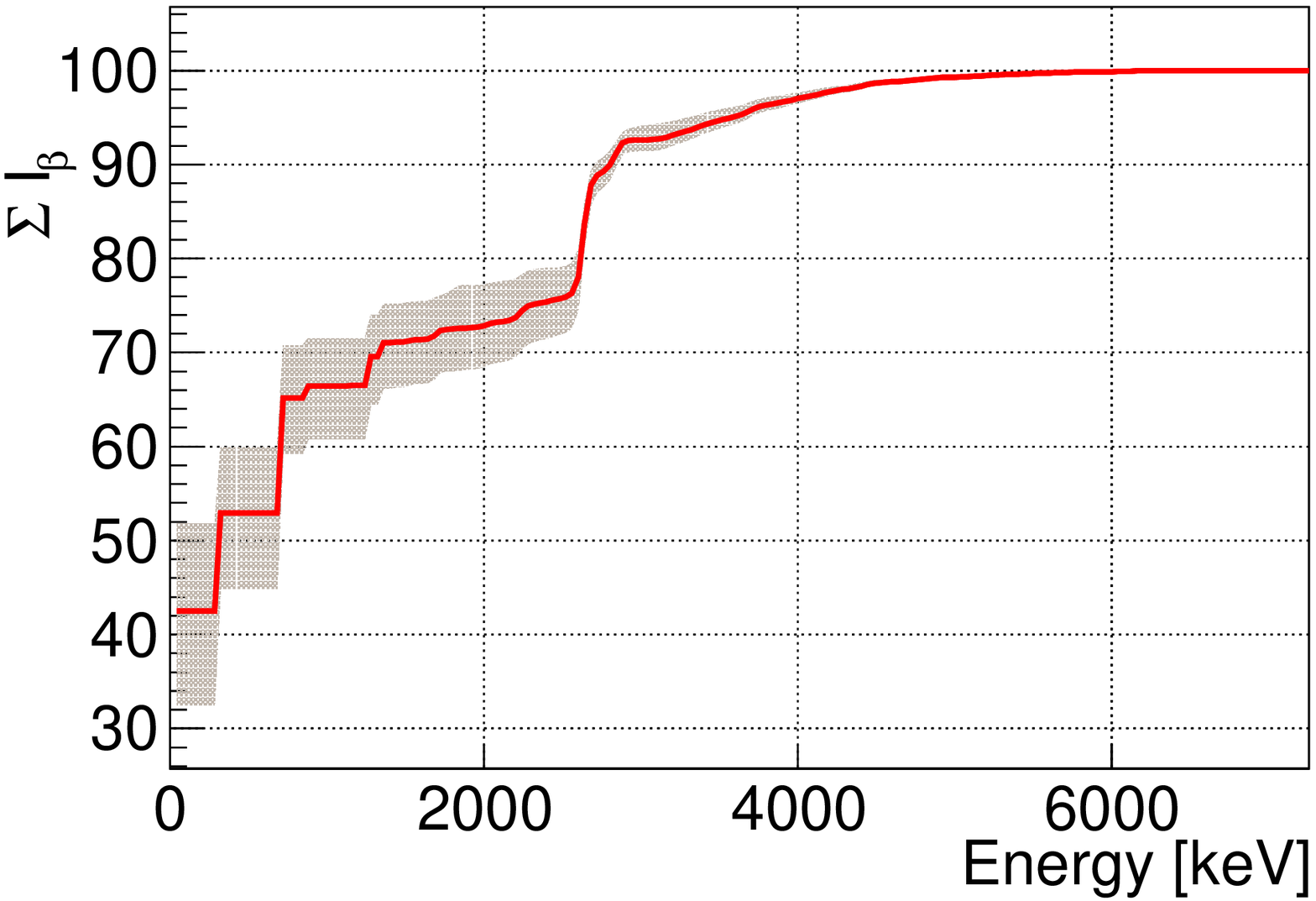} 
\caption{Same caption as in Fig.~\ref{100gsNb_fit} but for $^{102\text{m}}$Nb. In this case there is no previous data from ENSDF to compare with.}
\label{102mNb_fit}
\end{center}
\end{figure}

The quality of the reproduction of the experimental spectrum in the TAGS analysis is shown in Fig.~\ref{102mNb_fit} top, while the accumulated $\beta$ intensity distribution is presented in Fig.~\ref{102mNb_fit} bottom. No previous $\beta$-decay data were known for comparison. For the same reason it was not possible to optimize the branching ratio matrix by checking the $\gamma$-intensities. A large g.s. feeding intensity is obtained from the TAGS analysis: $42.5_{-10}^{+9}\%$. It was also calculated by means of a $\beta$-$\gamma$ counting method, as for $^{100\text{gs}}$Nb. Using this method a 43.5(24)$\%$ value was obtained, in agreement with the value from the TAGS analysis.

Systematic uncertainties included changes in the normalization factor of the contaminants. An acceptable reproduction of the experimental TAGS spectrum was still compatible with a change in the normalization of the summing-pileup up to 30$\%$, and with a 5$\%$ change for the contamination of the decay of $^{102}$Zr. Two possible spectral shapes discussed above for the contamination of the decay of $^{102}$Zr were included in the uncertainty estimate. Both come from our TAGS analysis, since the simulation with the information from ENSDF, corresponding to the spectral shape a) in Fig.~\ref{102Zr_TAGS}, did not result in a good subtraction. The effect of the first bin included in the analysis was also investigated, which gives the main contribution to the error budget of the g.s. feeding intensity.

\subsection{$^{102\text{gs}}$Nb}

The analysis of the decay of the high-spin ground state (4$^+$) was carried out permitting direct feeding to levels with spin-parity assignments $3^+$, $4^+$ and $5^+$ (allowed transitions). In the known part of the level scheme this means direct feeding only to levels at 743.7~keV (4$^+$) and at 1245.54~keV (3$^+$) excitation energy.  Similarly to the $^{100\text{m}}$Nb case, the contribution of the decay of the low-spin isomer has been considered as a contamination. This contribution has been normalized with the peak associated with the $0^+$ state at 698.26~keV, only populated in the decay of $^{102\text{m}}$Nb. As in the case of $^{100}$Nb, the shape of the subtracted spectrum at low energies, influenced by the penetration of $\beta$-particles, was used to verify the normalization factor.

\begin{figure}[h]
\begin{center}
\includegraphics[width=0.5 \textwidth]{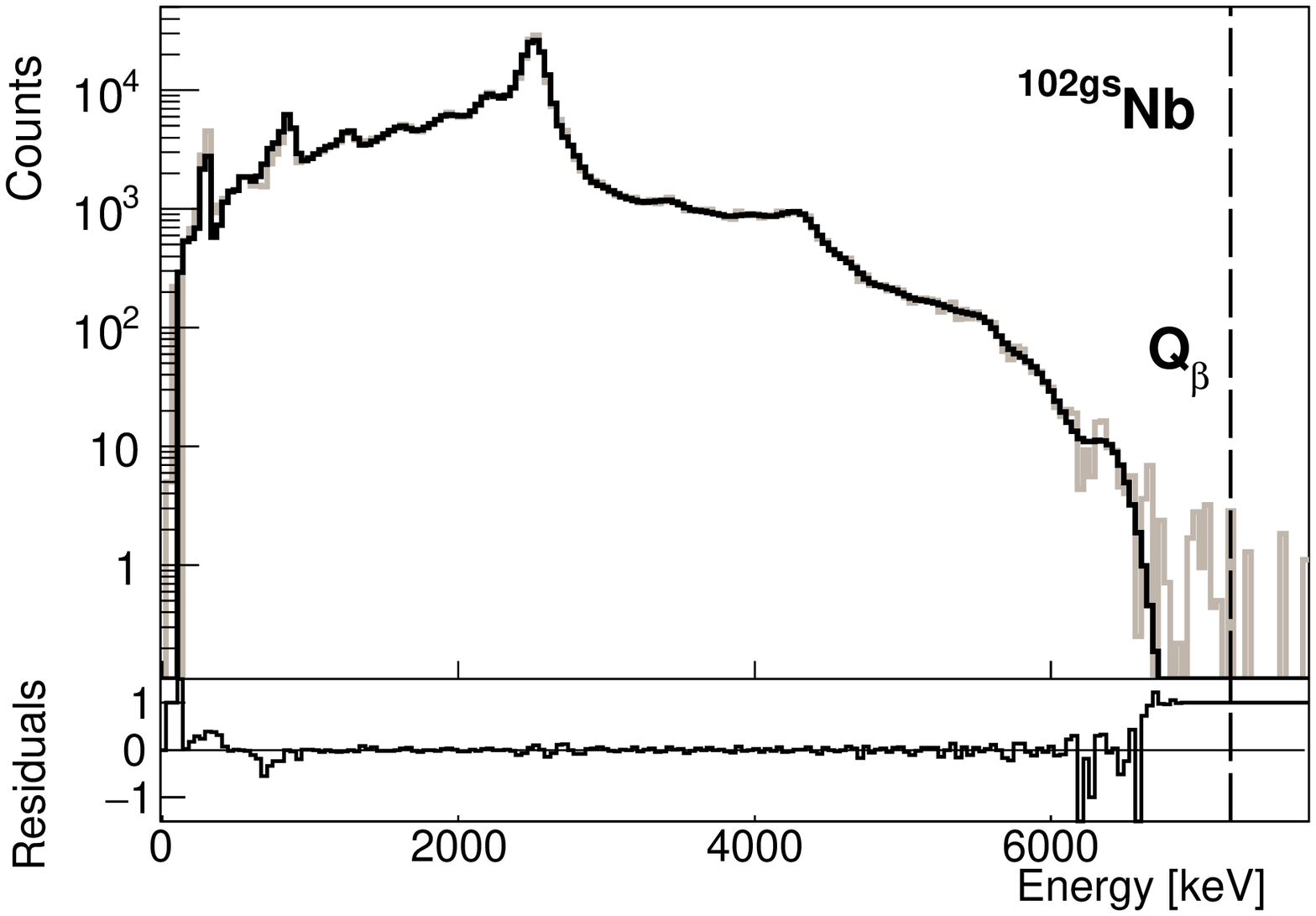}\\
\includegraphics[width=0.5 \textwidth]{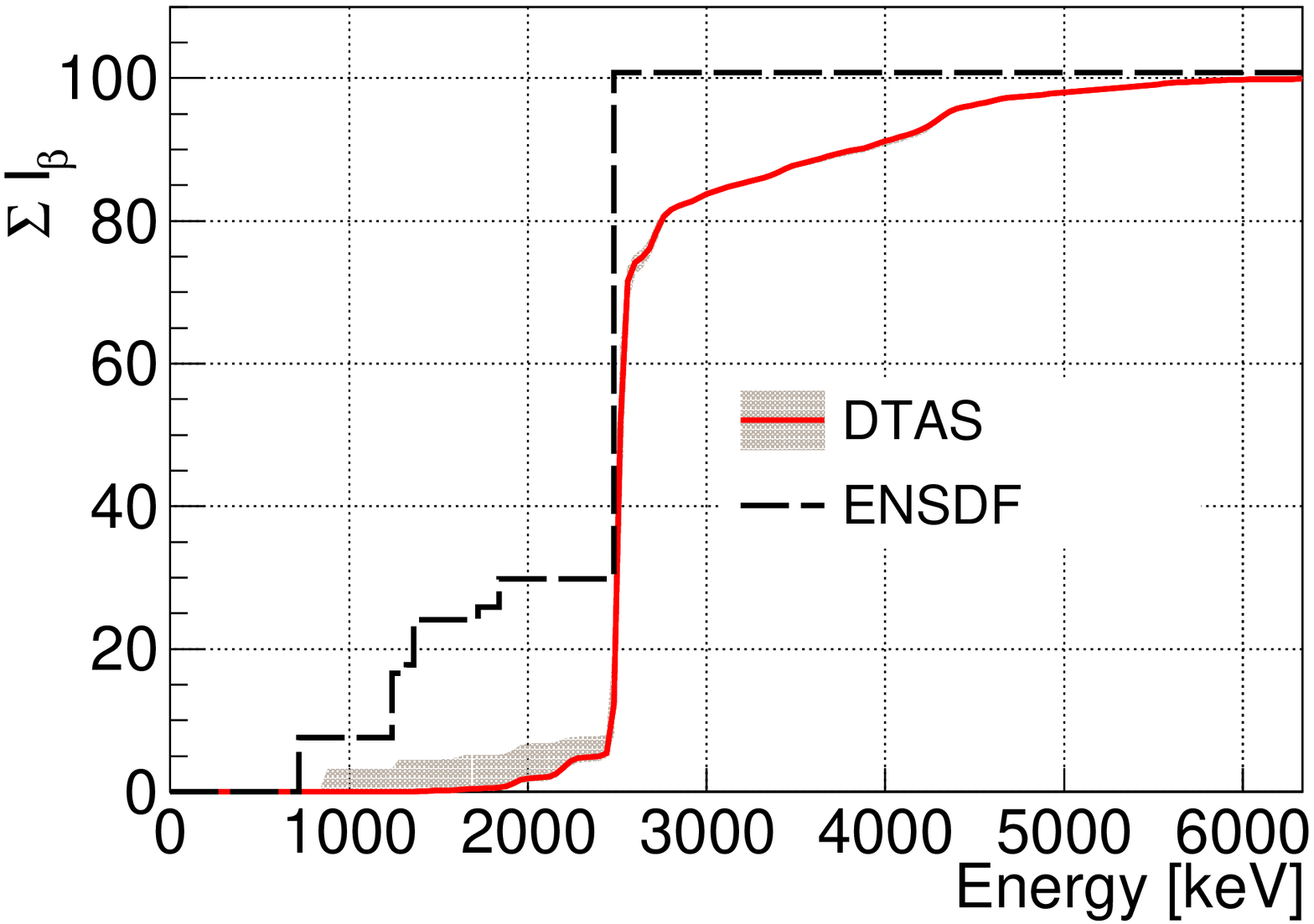} 
\caption{Same caption as in Fig.~\ref{100gsNb_fit} but for $^{102\text{gs}}$Nb.}
\label{102gsNb_fit}
\end{center}
\end{figure}

Similar to the case of $^{100\text{m}}$Nb, a branching ratio matrix modified to reproduce the known $\gamma$-intensities from high resolution experiments \cite{NDS_A102} notably improves the reproduction of the experimental TAGS spectrum. The $\gamma$-intensities with the original branching ratio matrix and the ones obtained with the modified branching ratio matrix are compared in Table \ref{tableIg102gsNb} with the high-resolution values. If the branching ratio matrix is not modified, the best reproduction of the experimental TAGS spectrum requires direct feeding to the 2$^+$ level at 847~keV. However, we have used the modified branching ratio matrix for the determination of the reference $\beta$-intensity distribution of the TAGS analysis. The quality of the reproduction of the experimental spectrum with the TAGS analysis is shown in Fig.~\ref{102gsNb_fit} top panel. In the bottom panel of Fig.~\ref{102gsNb_fit} we compare the TAGS accumulated $\beta$ intensity distribution with the previous data from ENSDF \cite{NDS_A102}, revealing a clear Pandemonium effect in previous measurements.

The reproduction of the experimental TAGS spectrum was compatible with a change in the normalization of the summing-pileup of up to 20$\%$, and with a change of up to 20$\%$ for the contamination of $^{102\text{m}}$Nb. In addition, as in the analysis of $^{100\text{m}}$Nb, we have considered the effect of subtracting the experimental low-spin spectrum (free of contaminants) or a simulated one. This allowed the study of the impact of the spectral shape of the $^{102}$Zr contamination in the analysis.

\begin{table}[h]
\centering
\begin{tabular}{|c|ccc|}
Energy [keV] & $I_{\gamma}$ ENSDF & $I_{\gamma}$ DTAS & $I_{\gamma}$ DTAS$^{*}$ \\ \hline
296    &	0.794	 &   0.881 & 0.790 \\
697    &	0.019  &   0.039 & 0.007 \\
743     &	0.196	 &   0.297 & 0.190 \\
847     &	0.505	 &   0.129 & 0.500 \\
1245    &	0.230  &   0.045 & 0.230 \\
1249     &	0.022  &   0.019 & 0.004 \\
1327    &	0.012 &   0.019 & 0.004 \\
1398    &	0.079 &   0.024 & 0.006 \\
\hline
\end{tabular}
\caption{\label{tableIg102gsNb} Absolute $\gamma$-intensities (per 100 decays) de-exciting the main levels populated in the decay of $^{102\text{gs}}$Nb $^{102}$Mo. The second column corresponds to the intensities obtained from high resolution data \cite{NDS_A102}. The third column gives the intensities obtained with DTAS for the base analysis, whereas the intensities obtained with a modified branching ratio matrix are presented in the fourth column (DTAS$^{*}$).}
\end{table}

\section{Cross-checks of the TAGS analyses \label{crosschecks}}

\begin{figure}[!hbt]
\begin{center}
\includegraphics[width=0.5 \textwidth]{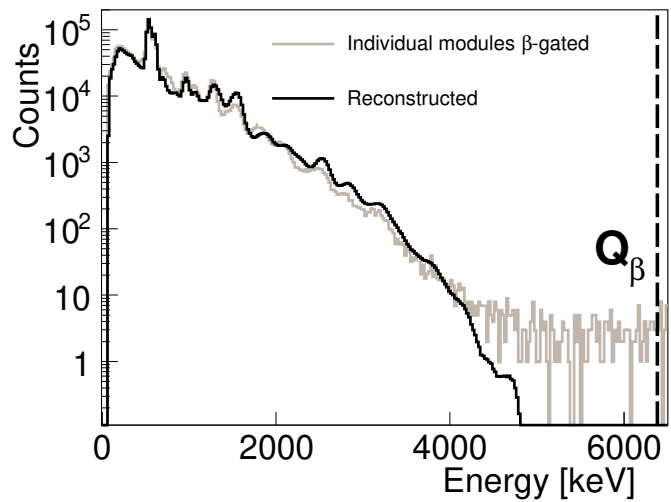} 
\caption{Experimental $\beta$-gated spectrum of the individual modules for $^{100\text{m}}$Nb free of contaminants (grey) and MC spectrum obtained using the results of the TAGS analysis (black).} 
\label{100mNb_individuals}
\end{center}
\end{figure}

The segmentation of DTAS allows one to check other experimental observables to validate the results of the analyses. One of them is the reproduction of the $\beta$-gated spectra of the individual modules, where DTAS can be considered as a conventional $\gamma$-spectroscopy array of detectors. A MC simulation using the DECAYGEN event generator \cite{TAS_decaygen} with the branching-ratio matrix and the final $\beta$-intensity distribution from each of our analyses as input, is used to compare with the experimental spectra. In Fig.~\ref{100mNb_individuals} we show the comparison with the experimental sum of individual modules free of contaminants for $^{100\text{m}}$Nb. Although it is clear that the agreement is not comparable to the one reached for the total spectrum (Fig. \ref{100mNb_fit} top) it can be considered acceptable. A similar situation is found for the other three cases studied.

\begin{figure*}[h]
\begin{center} 
\begin{tabular}{cc}
\includegraphics[width=0.5 \textwidth]{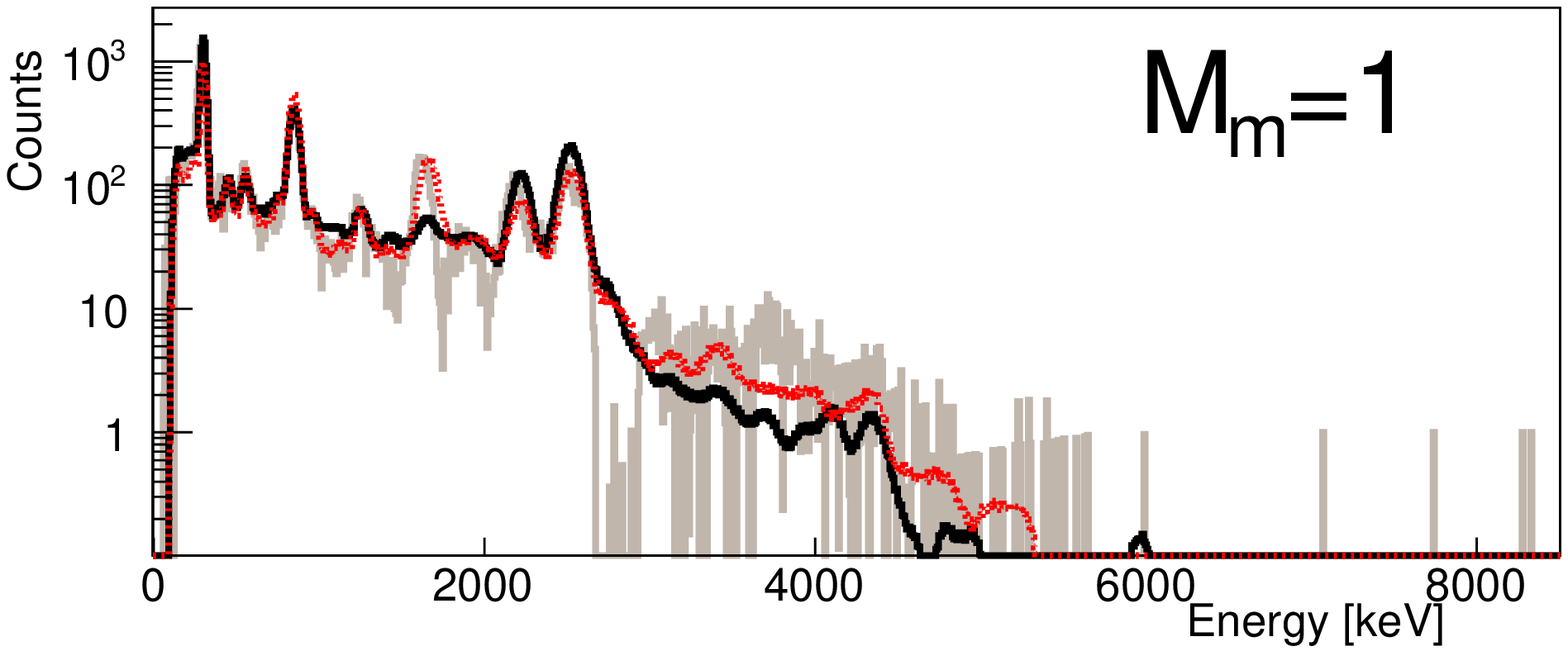} &
\includegraphics[width=0.5 \textwidth]{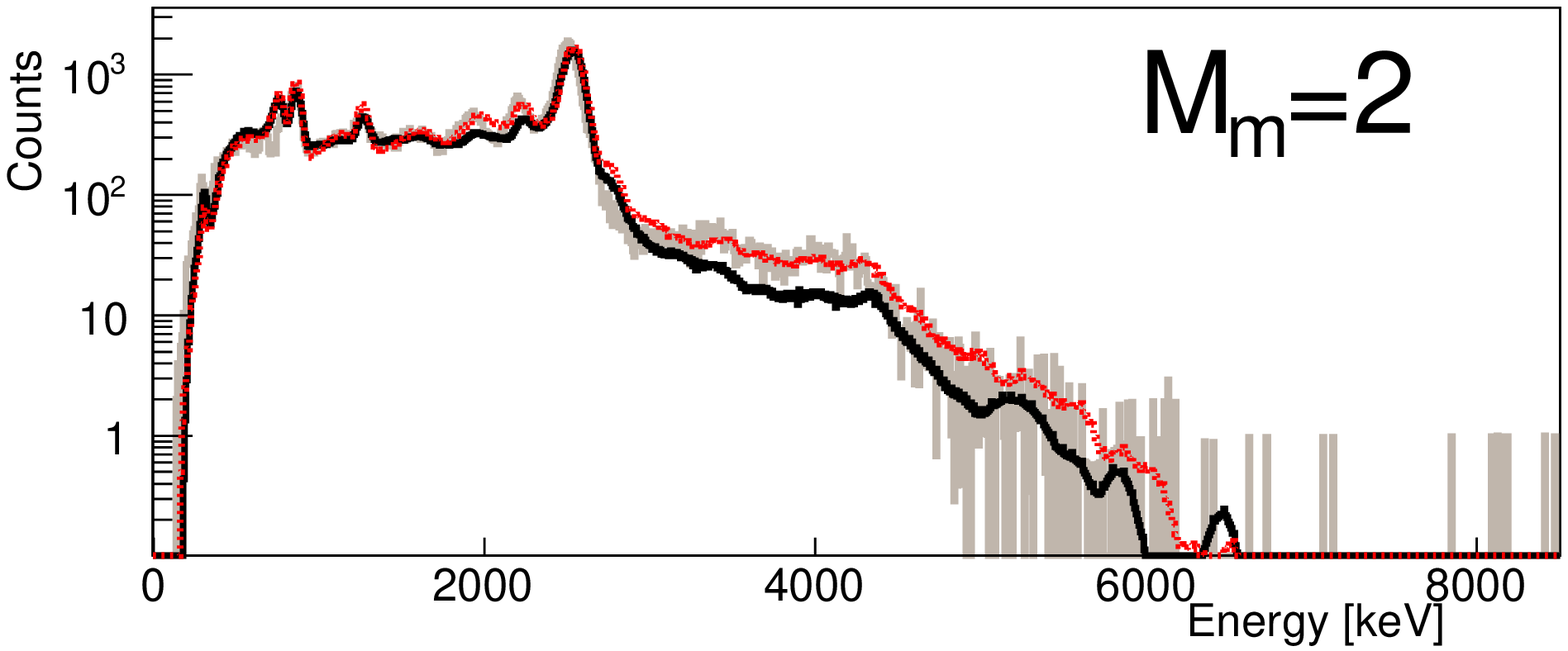} \\
\includegraphics[width=0.5 \textwidth]{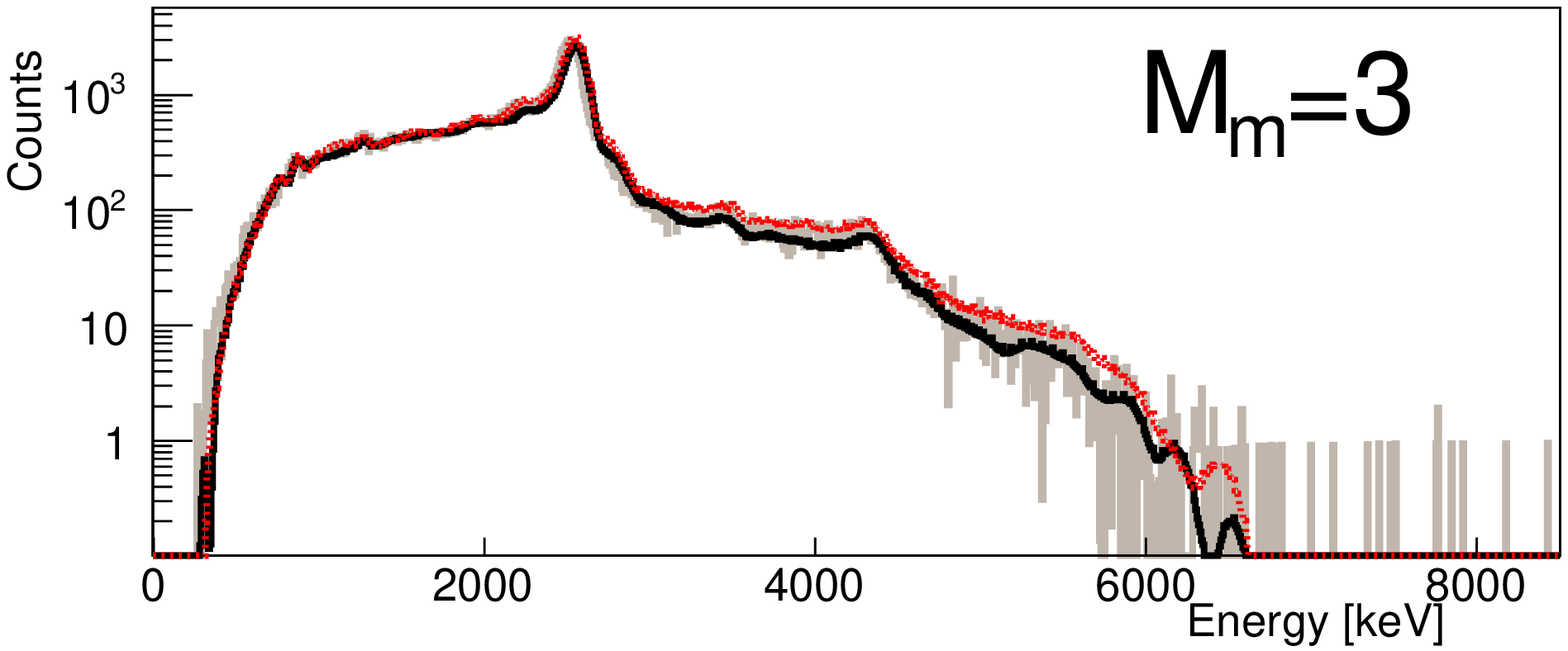} &
\includegraphics[width=0.5 \textwidth]{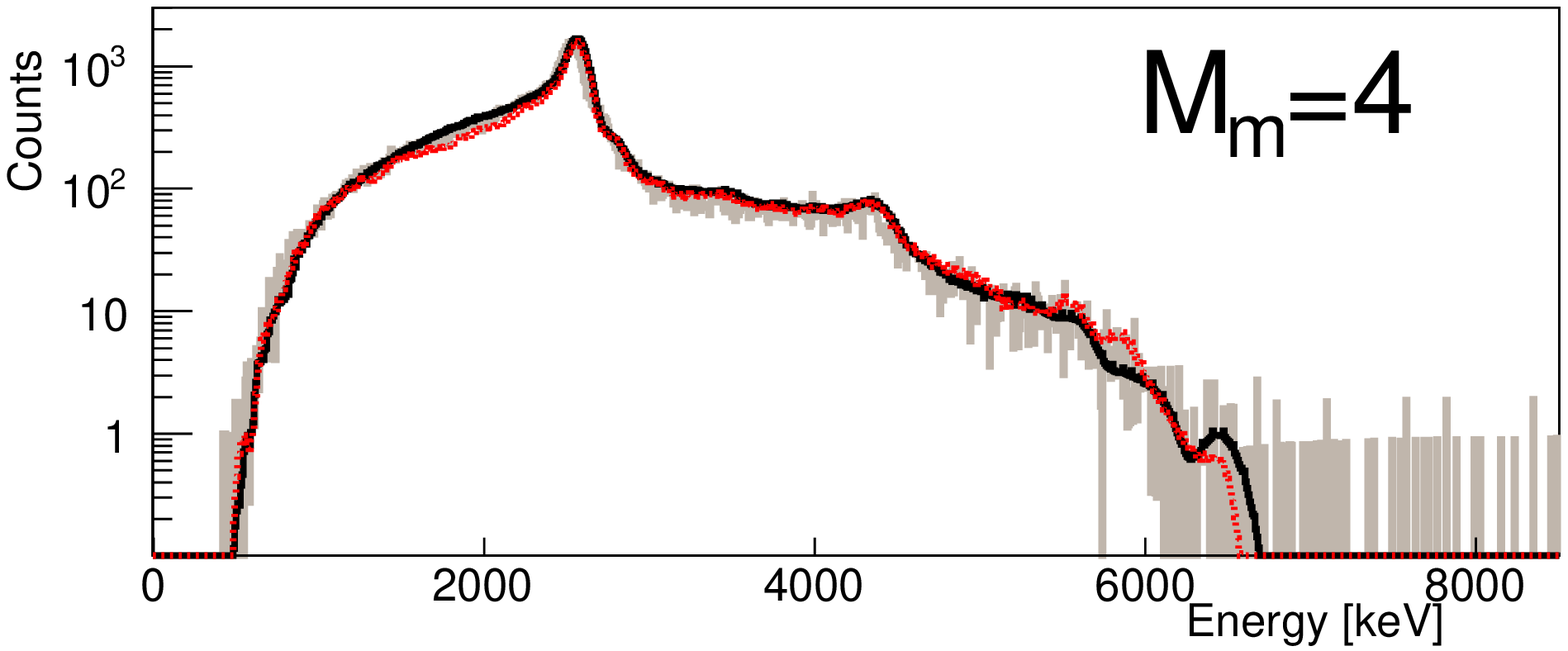} \\ 
\includegraphics[width=0.5 \textwidth]{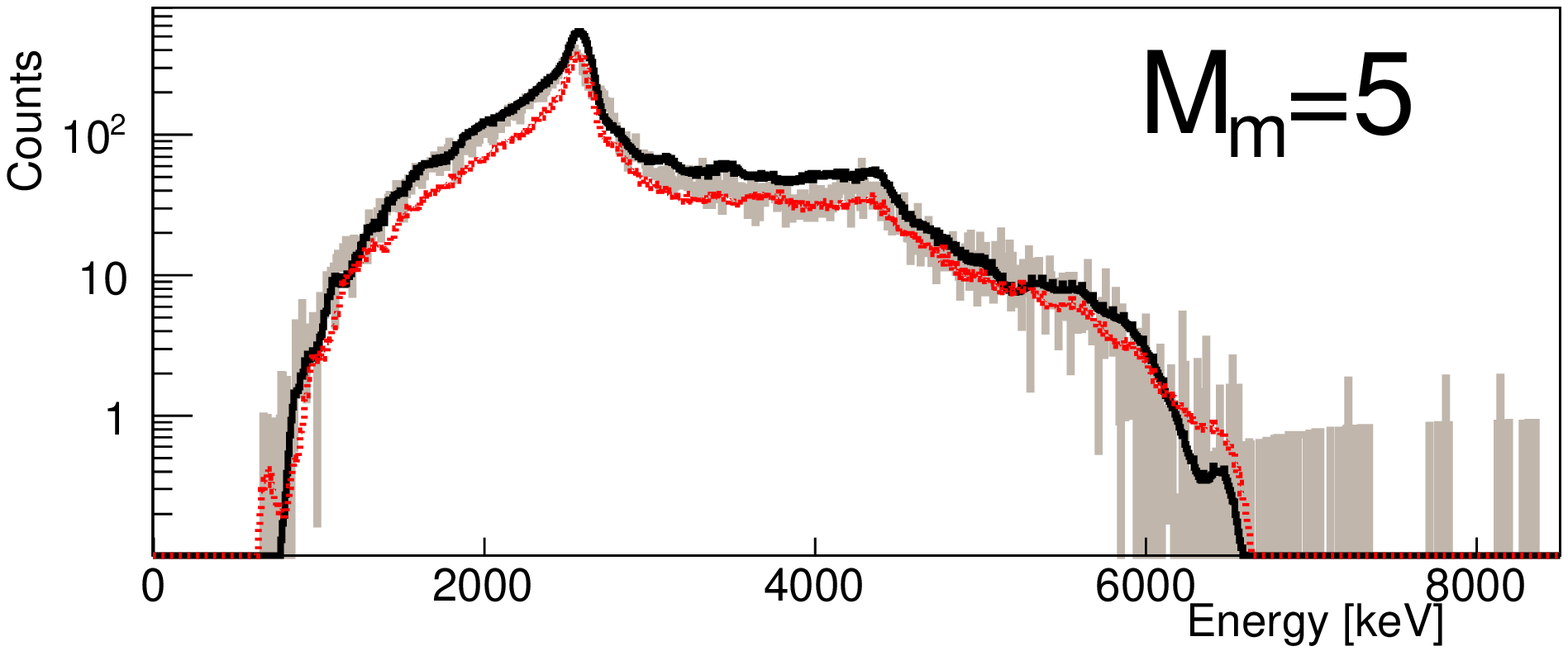} &
\includegraphics[width=0.5 \textwidth]{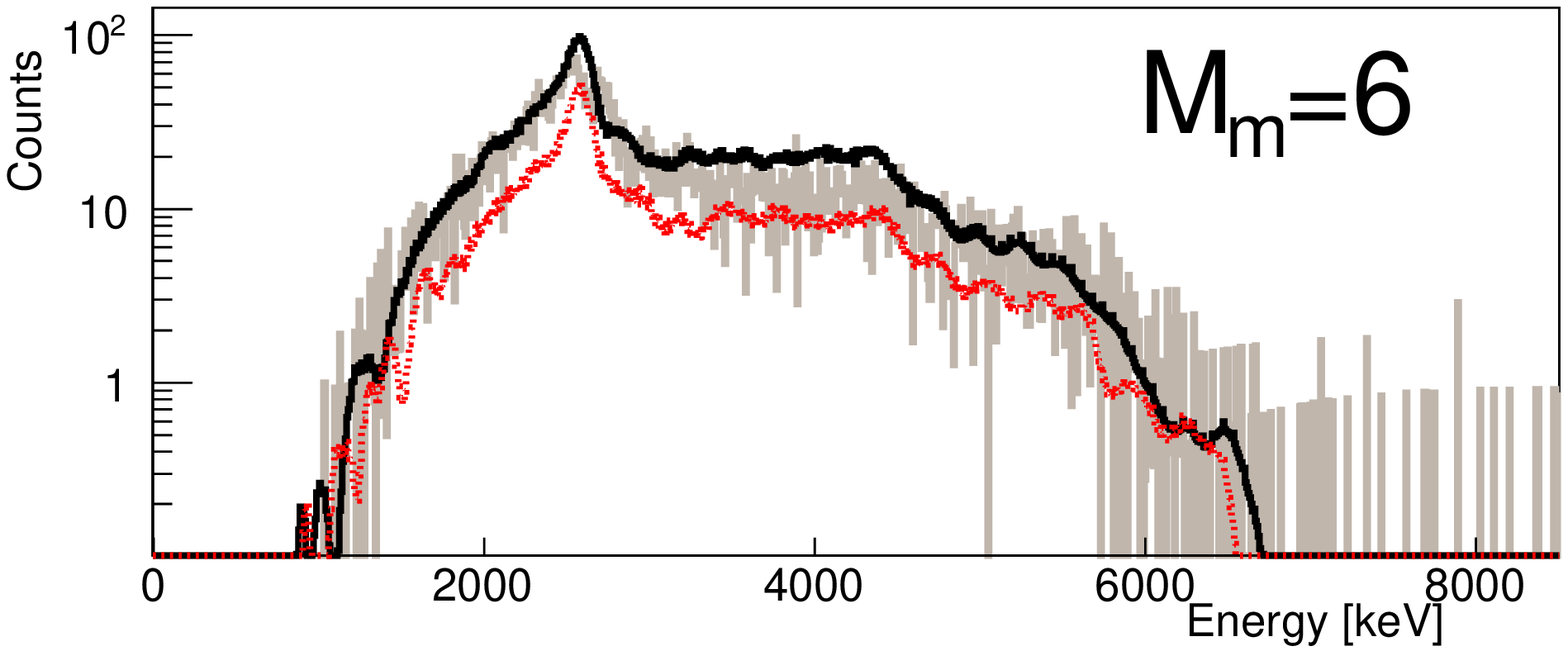} 
\end{tabular}
\caption{$^{102\text{gs}}$Nb experimental spectra after subtracting all the contaminants (solid grey) compared with the MC simulation using the original branching ratio matrix (solid black) and with the MC simulation after modifying the branching ratio matrix (dotted red). The sum-energy spectra with a condition on module multiplicity $M_m$ from 1 to 6 is shown.}
\label{MC_multiplicities}
\end{center}
\end{figure*}

The sensitivity of the individual spectra to the branching ratio matrix is however limited, because it reflects essentially the energy distribution of individual $\gamma$-rays, since more than two $\gamma$-rays are seldom detected in the same module. We can extract more information checking the $\beta$-gated TAGS spectra with conditions in the module-multiplicity ($M_m$), defined as the number of modules that fire above the threshold in one event. In the module-multiplicity gated spectra the $\gamma$-multiplicity of the de-excitation cascades of our branching ratio matrix plays a crucial role. In general a reasonable reproduction of the multiplicity-gated spectra was obtained. As an example, in Fig.~\ref{MC_multiplicities} the module-multiplicity spectra up to $M_m$=6 for $^{102\text{gs}}$Nb are shown. The experimental spectra free of contaminants are compared with the MC simulation using the DECAYGEN event generator. We show in this case the effect of modifying the branching ratio matrix to reproduce the $\gamma$-intensity of low energy transitions from high resolution experiments. As expected it improves the agreement for $M_m$=1 and 2 at low energies, but also at high energies. For $M_m$=3 and 4 the impact is very small. For $M_m$=5 some differences appear between low and high energies. Only for $M_m$=6 the spectrum appears to be somewhat smaller. Overall the agreement is quite good, demonstrating the quality of the analysis.

\section{Reactor summation calculations \label{reactor}}

The impact of the present results on reactor summation calculations was already discussed in \cite{PRL_Nb}, where special emphasis was placed on the antineutrino spectrum calculations. The present TAGS results showed a large impact on antineutrino spectrum summation calculations in the energy region of the shape distortion at 5-7~MeV. An increase of up to 2$\%$ for $^{235}$U and up to 6$\%$ for $^{239}$Pu was found, specially due to the new decay data for the A=102 niobium decays. The impact of the data reduces for the first time the discrepancy between the summation calculations and the measured antineutrino spectra in the region of the shape distortion. 

Here additional details about the reactor Decay Heat (DH) calculations are given. In the summation method for DH calculations an inventory of radioactive isotopes is first obtained by solving the linear system of coupled first order differential equations that describe the creation and decay of fission products in a nuclear reactor (see for example \cite{Nichols_report} for more details). With this inventory, the decay heat of fission products as a function of time can be computed by summing the energy released by the decay of each nucleus weighted by the activity at this time, according to Equation \eqref{decayheat_eq}:

\begin{equation}\label{decayheat_eq}
f(t) =\sum\limits_{i}(\overline{E}_{\beta,i} + \overline{E}_{\gamma,i})\lambda_i N _i(t) 
\end{equation}

\noindent where $f(t)$ is the power function, $\overline{E}_i$ is the mean decay energy of the $i$th nuclide ($\beta$ and $\gamma$), $\lambda_i$ is the decay constant of the $i$th nuclide, and $N_i(t)$ is the number of nuclei $i$ at cooling time $t$. The mean $\gamma$ and $\beta$ energies are defined as:

\begin{equation}\label{mean_energies_eq}
\begin{aligned}
\overline{E}_{\gamma}=\sum_{i}I_{\beta}(E_i)E_i  
&&
,
&& 
\overline{E}_{\beta}=\sum_{i}I_{\beta}(E_i)<E_{\beta i}>
\end{aligned}
\end{equation}

\noindent where $I_{\beta}(E_i)$ is the probability of $\beta$ feeding to level $E_i$ and $\langle E_{\beta i}\rangle$ is the mean energy of the $\beta$ particles emitted when level $i$ is fed.

The $\beta$-intensity distributions obtained in this work have been used to evaluate the average $\gamma$ and $\beta$ energies. They were computed using Eqs. \eqref{mean_energies_eq}, where $\langle E_{\beta i}\rangle$ was calculated employing subroutines from the $\log ft$ program of NNDC \cite{logftNNDC} assuming an allowed $\beta$ shape. The resulting average energies obtained in this work are listed in Table \ref{MeanE}, where the quoted uncertainties are deduced from the evaluation of the average energies for all the space of solutions obtained for each case. The values from the reference databases ENDF/B-VII.1 and JEFF-3.1.1 are included in Table \ref{MeanE} for comparison. The effect of the Pandemonium systematic error is observed in the data available in the databases, as found for previous cases \cite{DecayHeat,vTAS_PRC,Simon_PRC}: the mean $\gamma$ energies are underestimated, while the mean $\beta$ energies are overestimated.

\begin{table*}[h]
\begin{center}
\begin{tabular}{c|c|c|c|c|c|c}
\hline \hline
  Decay & \multicolumn{3}{c|}{$\overline{E}_{\gamma}$ [keV]} & \multicolumn{3}{c}{$\overline{E}_{\beta}$ [keV]}  \\ 
  & TAGS & ENDF & JEFF & TAGS & ENDF & JEFF  \\  \hline
  $^{100\text{gs}}$Nb & 959(275) & 708(37) & 708 & 2414(133) & 2539(213) & 2484(209) \\
  $^{100\text{m}}$Nb & 2763(27) & 2213(69) & 2056 & 1706(13) & 1999(198) & 2039 \\
  $^{102\text{gs}}$Nb & 2764(57) & 2094(97) & 2094 & 1948(27) & 2300(169) & 2276(169) \\
  $^{102\text{m}}$Nb & 1023(170) & - & - & 2829(82) & - & - \\
  \hline \hline
\end{tabular}
\caption{\label{MeanE} Average $\gamma$ and $\beta$ energies of the decays studied used in the decay heat summation calculations. The present TAGS results are compared with the values available in the ENDF/B-VII.1 and JEFF-3.1.1 databases.}
\end{center}
\end{table*}

\begin{figure*}[h]
\begin{center} 
\includegraphics[width=0.8 \textwidth]{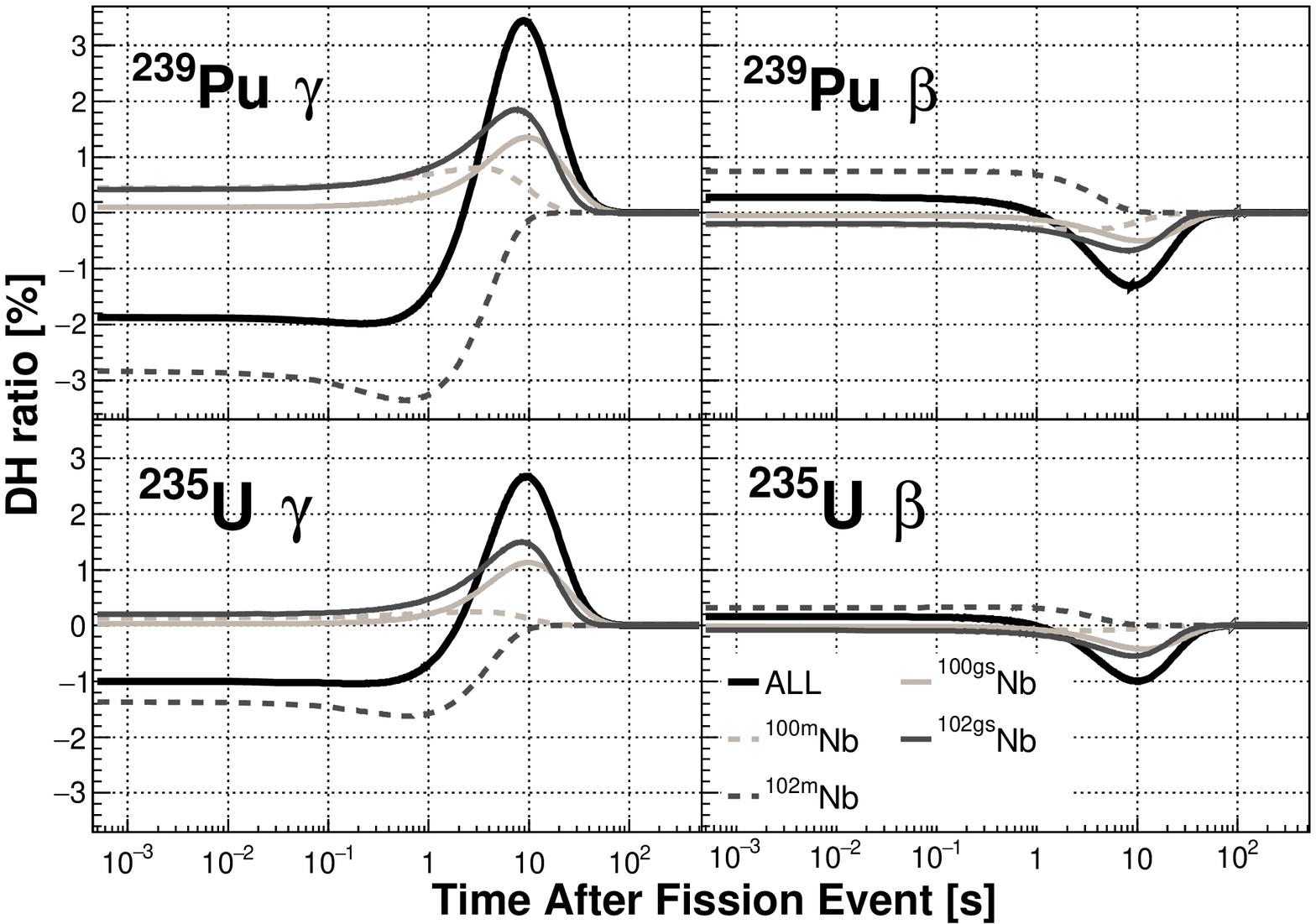} 
\caption{Ratio in $\%$ between the reactor decay-heat as a function of time including the data obtained in the present work and the reactor decay-heat calculated with the previous knowledge in the ENDF/B-VII.1 database. The $\gamma$ and $\beta$ components of the decay-heat for $^{235}$U and $^{239}$Pu are shown.}
\label{DH}
\end{center}
\end{figure*}

The impact of our new TAGS data on DH summation calculations has been evaluated using the ENDF/B-VII.1 database as a reference. For $^{102\text{m}}$Nb we compare with the simple estimate $\overline{E}_{\gamma}$=$\overline{E}_{\beta}$=$Q_{\beta}/3$, provided the lack of previous information for this decay. In Fig.~\ref{DH} we present the ratio between the DH calculation with/without the new TAGS data for $^{235}$U and $^{239}$Pu fissile isotopes. The largest effect is observed for both $\gamma$ components, specially due to the new data for $^{100\text{gs}}$Nb and $^{102\text{gs}}$Nb at 10~s (reaching values of about +3$\%$), while $^{102\text{m}}$Nb impacts notably at shorter times. The total impact in the $\beta$ component of $^{235}$U and $^{239}$Pu is particularly noticeable at 10~s (about -1$\%$).

\section{Conclusions \label{conclusions}}

In this work we have obtained the $\beta$ intensity distributions of the decays of $^{100\text{gs},100\text{m}}$Nb and $^{102\text{gs},102\text{m}}$Nb free from the Pandemonium effect by means of the TAGS technique. Even though these decays were in the priority lists of the International Atomic Energy Agency (IAEA) to improve decay heat and antineutrino spectrum summation calculations, the difficulties in producing and disentangling the decays of the isomers had remained as a big challenge until now. Thanks to the capabilities of IGISOL and JYFLTRAP, the well characterized DTAS detector, and the strategies followed to separate the different contributions, we were able to distinguish experimentally the decay of each pair of isomers. 

In the TAGS analyses of the decays of $^{100\text{gs},100\text{m}}$Nb and $^{102\text{gs}}$Nb we found previously undetected $\beta$-intensity at high excitation energy, while the $\beta$-intensity distribution of the decay of $^{102\text{m}}$Nb was obtained for the first time. A careful evaluation of the systematic uncertainties was performed, and we have also evaluated the quality of the analysis using the information of the individual module spectra and of the module-multiplicity gated spectra. A reasonable reproduction of these two additional observables is found when we use the results of the TAGS analyses as input in MC simulations to compare with the experiment. We have also modified the branching ratio matrices to reproduce the known $\gamma$-intensities at low excitation energies, in the cases where they are considered reliable. 

The impact of the present results on reactor summation calculations was studied. After reporting a large impact on antineutrino spectrum calculations \cite{PRL_Nb}, here we showed a notable impact on decay heat calculations. A global increase of around 3$\%$ at 10~s in the $\gamma$ component of $^{235}$U and $^{239}$Pu is found, while at shorter times a decrease of 1$\%$ and 3$\%$ is observed for $^{235}$U and $^{239}$Pu, respectively. The total impact in the $\beta$ components of both fissile isotopes represents a reduction of 1$\%$ at 10~s. 

In addition, in this work we discussed a method to extract the decays of the zirconium parents that were contaminants in the measurements of the low-spin niobium isomers. An analysis of these decays allowed us to compare the results of the $\beta$ strength with QRPA calculations, pointing to a dominance of prolate configurations in the ground states for these nuclei. 

Finally, the methods applied in this work will be of great interest for the study of other systems with isomers measured in the same experimental campaign and for those planned for future experiments.


\begin{acknowledgments}
This work has been supported by the Spanish Ministerio de Econom\'ia y Competitividad under Grants No. FPA2011-24553, No. AIC-A-2011-0696, No. FPA2014-52823-C2-1-P, No. FPA2015-65035-P, No. FPI/BES-2014-068222, No. FPA2017-83946-C2-1-P and the program Severo Ochoa (SEV-2014-0398), by the Spanish Ministerio de Educaci\'on under the FPU12/01527 Grant, by the European Commission under the FP7/EURATOM contract 605203 and the FP7/ENSAR contract 262010, and by the $Junta~para~la~Ampliaci\acute{o}n~de~Estudios$ Programme (CSIC JAE-Doc contract) co-financed by FSE. We acknowledge the support of the UK Science and Technology Facilities Council (STFC) Grant No. ST/P005314/1. This work
was also supported by the Academy of Finland under the Finnish Centre of Excellence Programme (Project No. 213503, Nuclear and Accelerator-Based Physics Research at JYFL). The authors thank the IAEA for supporting and encouraging the work in this field.
\end{acknowledgments}

\bibliography{Nb}

\end{document}